\documentclass[usenatbib]{mn2e}
\usepackage{epsfig,aastex_hack, amssymb, plotmore,amsmath}

\newcommand{\mincir}{\raise
-2.truept\hbox{\rlap{\hbox{$\sim$}}\raise5.truept 
\hbox{$<$}\ }}
\newcommand{\magcir}{\raise
-2.truept\hbox{\rlap{\hbox{$\sim$}}\raise5.truept
\hbox{$>$}\ }}
\newcommand{\minmag}{\raise-2.truept\hbox{\rlap{\hbox{$<$}}\raise
6.truept\hbox
{$>$}\ }}

\newcommand{\beq}{\begin{equation}}
\newcommand{\eeq}{\end{equation}}

\newcommand{\be}{\begin{equation}}
\newcommand{\ee}{\end{equation}}
\newcommand{\ba}{\begin{eqnarray}}
\newcommand{\ea}{\end{eqnarray}}
\newcommand{\brr}{\begin{array}}
 
\newcommand{\err}{\end{array}}
\newcommand{\bc}{\begin{center}}
\newcommand{\ec}{\end{center}}

\newif\ifAMStwofonts
\AMStwofontstrue

\def\lsim{\mathrel{\hbox{\rlap{\hbox{\lower4pt\hbox{$\sim$}}}\hbox{$<$}}}}
\def\gsim{\mathrel{\hbox{\rlap{\hbox{\lower4pt\hbox{$\sim$}}}\hbox{$>$}}}}

\def\LCDM{$\Lambda$CDM }

\newcommand{\rhomean}{\overline{\rho}_{\rm m}}

\newcommand{\kpch}{\>h^{-1} {\rm kpc}}

\newcommand{\Mpch}{\>h^{-1} {\rm Mpc}}

\newcommand{\Msolh}{\>h^{-1} M_{\odot}}

\newcommand{\xihm}{\xi_{\rm hm}(r)}

\newcommand{\xioneh}{\xi_{\rm 1h}(r)}
\newcommand{\xitwoh}{\xi_{\rm 2h}(r)}

\DeclareMathAlphabet{\mathsc}{OT1}{cmr}{m}{sc}
\def\testbx{bx}%
\DeclareRobustCommand{\ion}[2]{%
\relax\ifmmode
\ifx\testbx\f@series
{\mathbf{#1\,\mathsc{#2}}}\else
{\mathrm{#1\,\mathsc{#2}}}\fi
\else\textup{#1\,{\mdseries\textsc{#2}}}%
\fi}

\title[Halo-mass and galaxy-mass cross-correlations]
{Understanding the shape of the halo-mass and galaxy-mass cross-correlation 
functions}
\author[E. Hayashi and S.D.M. White] {Eric Hayashi
and Simon D.M. White\\Max Planck Institute for
Astrophysics,  Karl-Schwarzschild Strasse 1,  D-85748 Garching, Germany}

\begin{document}

\maketitle
\begin{abstract}
We use the Millennium Simulation to measure the cross-correlation between halo
centres and mass (or equivalently the average density profiles of dark haloes)
in a $\Lambda$CDM cosmology. We present results for radii in the range $10~
h^{-1}{\rm kpc} < r < 30~h^{-1}{\rm Mpc}$ for halo masses in the range
$4\times 10^{10}~h^{-1}{\rm M_\odot} < M_{200} < 4\times 10^{14}~h^{-1}{\rm
M_\odot}$. Both at $z=0$ and at $z=0.76$ these cross-correlations are
surprisingly well fit by approximating the inner region by a density profile
of NFW or Einasto form, the outer region by a biased version of the linear
mass autocorrelation function, and by adopting the maximum of the two where
they are comparable. We use a simulation of the formation of galaxies within
the Millennium Simulation to explore how these results are reflected in
cross-correlations between galaxies and mass. These are directly observable
through galaxy-galaxy lensing.  Here also we find that simple models can
represent the simulation results remarkably well, typically to $\lsim
10\%$. Such models can be used to extend our results to other redshifts, to
cosmologies with other parameters, and to other assumptions about how galaxies
populate dark haloes.  The characteristic features predicted in the
galaxy-galaxy lensing signal should provide a strong test of the $\Lambda$CDM
cosmology as well as a route to understanding how galaxies form within it.
\end{abstract}


\section{Introduction}
\label{sec:intro}

Weak gravitational lensing has opened a new window onto the large scale
distribution of matter.  Gravitational lensing by foreground mass induces
correlated distortions, or shear, in the observed shapes of distant galaxies.
In galaxy-galaxy lensing, the signal from many galaxies is added together in
order to measure the average (projected) distribution of mass around galaxies.
This can be interpreted as the mass in the extended dark matter halos which
surround galaxies, or, more generally, as the cross-correlation between lens
galaxies and the projected mass distribution.  Several groups have
successfully applied this technique to large imaging surveys to derive
constraints on the mass associated with galaxies as a function of galaxy
properties such as luminosity and morphology
\citep{BRAINERD96,DELLANTONIO96,GRIFFITHS96, HUDSON98, FISCHER00,
MCKAY01,HOEKSTRA03, SHELDON04}.

Theoretical predictions for cross-correlations between galaxies and mass,
$\xi_{\rm gm}$, have made use both of numerical simulations
\citep{GUZIK01,YANG03, TASITSIOMI04, WEINBERG04} and of analytic halo models
\citep{SELJAK00,GUZIK02}.  \cite{TASITSIOMI04} show that the amplitude and
shape of $\xi_{\rm gm}$ predicted by cosmological simulations, when combined
with a simple model for populating halos with galaxies, are in good agreement
with observational results based on Sloan Digital Sky Survey (SDSS) data
\citep{SHELDON04}.  \cite{MANDELBAUM05} also find that these simulation
results can be accurately reproduced by the halo model of \cite{SELJAK00} and
\cite{GUZIK02}.

In this work we calculate the cross-correlation between halos and mass,
$\xi_{\rm hm}$, and between galaxies and mass, $\xi_{\rm gm}$, in the
Millennium Simulation, a very large, high-resolution simulation of a \LCDM
cosmology.  We also present simple models for $\xi_{\rm hm}$ and $\xi_{\rm
gm}$ which can be used to interpret the shapes of these functions and to
extend our results to other redshifts, cosmologies and halo population
models.

The organization of this paper is as follows.  In \S\ref{sec:sim} we describe
the Millennium Simulation and the halo and galaxy catalogues used in this
study.  In \S\ref{sec:xihm} we show $\xi_{\rm hm}$ calculated for halo samples
with a range of masses and we present a model which accurately reproduces
these results. In \S\ref{sec:xigm} we show $\xi_{\rm gm}$ for both central and
satellite galaxies as a function of their luminosity, and we present models
for each of these and for the combined galaxy sample.  Finally, we summarize
our results in \S\ref{sec:summary}.

\section{Simulations}
\label{sec:sim}

This study makes use of the Millennium
Simulation\footnote{http://www.mpa-garching.mpg.de/millennium/}
\citep{SPRINGEL05}, a large cosmological N-body simulation carried out by the
Virgo Consortium\footnote{http://www.virgo.dur.ac.uk/}.  In this simulation a
flat \LCDM cosmology is adopted, with $\Omega_{\rm dm}=0.205$, $\Omega_{\rm
b}=0.045$ for the current densities in cold dark matter and baryons, $h=0.73$
for the present dimensionless value of the Hubble constant, $\sigma_8=0.9$ for
the {\it rms} linear mass fluctuation in a sphere of radius $8~\Mpch$
extrapolated to $z=0$, and $n=1$ for the slope of the primordial fluctuation
spectrum.  The simulation follows $2160^3$ dark matter particles from $z=127$
to $z=0$ within a periodic cube $L_{\rm box}=500~\Mpch$ on a side.  The
individual particle mass is thus $8.6~\times 10^8~\Msolh$, and the
gravitational force is softened with a Plummer-equivalent comoving softening
of $5~\kpch$. Initial conditions were generated using the Boltzmann code
CMBFAST \citep{SELJAK96} to generate a realization of the desired power
spectrum which was then imposed on a glass-like uniform particle load
\citep{WHITE96}. A modified version of the TREE-PM N-body code GADGET2
\citep{SPRINGEL01,SPRINGEL05} was used to carry out the simulation and full
particle data are stored at 64 output times approximately equally spaced in
the logarithm of the expansion factor at early times and at roughly 300 Myr
intervals after $z=2$.

In each output of the simulation, halos are identified using a
friends-of-friends (FoF) groupfinder with a linking length of $b=0.2$
\citep{DAVIS85}.  Each FoF halo is decomposed into a collection of locally
overdense, self-bound substructures (or subhalos) using the SUBFIND algorithm
of \cite{SPR01}.  Of these subhalos, one is typically much larger than the
others and contains most of the mass of the halo.  We identify this as the
main subhalo and define its centre as the position of the particle with the
minimum potential.  The virial radius, $r_{200}$, is defined as the radius of
a sphere that encompasses a mean density 200 times the critical value, and the
virial mass, $M_{200}$, is the mass within this radius.

Semi-analytic techniques have been used to simulate the evolution of the
galaxy population within the Millennium Simulation, as described by
\cite{SPRINGEL05} and \cite{CROTON06}.  In this approach the evolution of the
baryonic component is followed using a set of simple prescriptions for gas
cooling, star formation, supernova and AGN feedback, chemical enrichment,
galaxy merging and other relevant physical processes.  These models can be
applied repeatedly to the stored histories of the dark matter halos and
subhalos with different parameter choices for the model, or indeed with
different physical assumptions in the modelling.  In this work we use the
semi-analytic galaxy catalogue described in detail in \cite{CROTON06}.  We
note that various aspects of the clustering properties of halos and galaxies
in the Millennium Simulation have been investigated in \cite{SPRINGEL05},
\cite{GAO05}, \cite{HARKER06},\cite{WANG06}, \cite{SPRINGEL06},
\cite{CROTON06}, \cite{LI06}, \cite{CROTON07} and \cite{GAO07}. Studies of
halo density profiles in the simulation have been carried out by \cite{NETO07}
and \cite{GAO07b}.

\section{Halo-mass cross-correlations}
\label{sec:xihm}
Given a density field, $\rho({\bf x})$, the density fluctuation field is
defined as
\begin{equation}
\delta({\bf x})=\frac{\rho({\bf x}) - \overline{\rho}}{\overline{\rho}}.
\label{eq:deltarho}
\end{equation}
The two-point autocorrelation function is defined as 
\begin{equation}
\xi({\bf r}) \equiv \langle \delta({\bf x)} \delta({\bf x} + {\bf r}) \rangle.
\label{eq:xidef}
\end{equation}
Assuming isotropy reduces this to a function of the separation, $\xi(r)$.
This is interpreted as a measurement of the excess probability above random of
finding a pair of objects with separation $r$ \citep{PEEBLES80}.  In the case
of two different populations of objects, $\delta_{\rm a}({\bf x})$ and
$\delta_{\rm b}({\bf x})$, the two-point cross-correlation function is given
by
\begin{equation}
\xi_{\rm ab}({\bf r}) \equiv \langle \delta_{\rm a}({\bf x)} \delta_{\rm b}({\bf x} + {\bf r}) \rangle.
\label{eq:xi2def}
\end{equation}
If we consider the cross-correlation between halo centres and mass, the
cross-correlation function $\xi_{\rm hm}(r)$ simply reflects the
spherically-averaged halo density profile averaged over all halos in the
sample.  This can be seen by combining eqs.~\ref{eq:deltarho} and
\ref{eq:xi2def} to give
\begin{equation}
\xihm = \frac{\langle \rho(r) \rangle - \rhomean}{\rhomean},
\end{equation}
where $r$ is the radial distance from the halo centre and $\rhomean =
\rho_{\rm crit} \Omega_{\rm m}$ is the mean density of the Universe.

\begin{figure}
\plotone{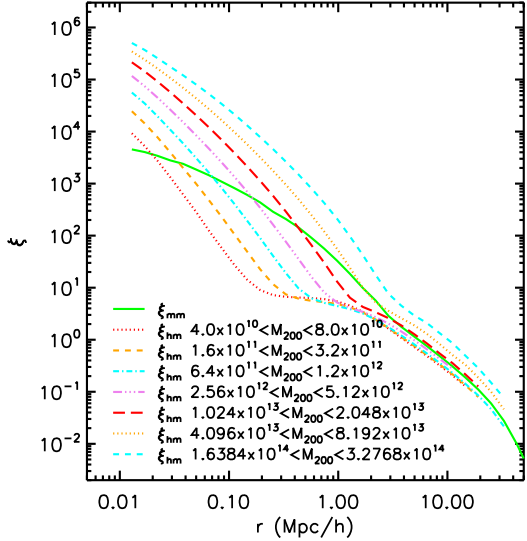}
\caption{Cross-correlations between halo centres and the mass in the
Millennium Simulation at $z=0$.  The solid curve shows the mass
autocorrelation function, $\xi_{\rm mm}$ for comparison.  On large scales, $r
\gsim 3~\Mpch$, the cross-correlation functions follow the mass
autocorrelation function with a constant bias factor.  On small scales,
$\xi_{\rm hm}(r)$ reflects the mean density profile of the individual
halos. Note the sharp transition between the two regimes.}
\label{fig:xigm_mlink}
\end{figure}

Figure~\ref{fig:xigm_mlink} shows halo-mass cross-correlations along with the
mass autocorrelation function for the Millennium Simulation.  The halo-mass
cross-correlations are computed using the centres of the main subhalos of
all FoF halos with mass $M_{200} \geq 4\times 10^{10}~\Msolh$,
corresponding to $\gsim 50$ particles.  Seven halo samples are shown, each
spanning a factor of two in mass.  The number of halos in each sample is
listed in Table~\ref{tab:halos}. We have computed cross-correlations also for
the mass ranges not included in this plot and table, and we have checked that
they also fit the models we discuss below. However, for clarity we refrain
from showing them or discussing them further.

\begin{table*}
\centering
\caption{Properties of halo samples taken from the Millennium Simulation.}
\begin{tabular}{llll|llll}
\hline
Mass range  & $N_{\rm halos}$ & $<M_{200}>$  &$<R_{200}>$ &Fit
$M_{200}$ &Fit $R_{200}$ & Fit $c_{200}$ & $r_{\rm trans}$ \\
$[\Msolh]$ & & $[\Msolh]$ & $\Mpch$ &  $[\Msolh]$ & $\Mpch$ & & $\Mpch$
\\ \hline
$[4.0,8.0]\times 10^{10}$ & 2491565 & $5.59\times10^{10}$ & 0.062 & $5.66\times10^{10}$ & 0.062 & 9.8 & 0.11 \\
$[1.6,3.2]\times 10^{11}$ & 774837 & $2.23\times 10^{11}$ & 0.098 & $2.26\times 10^{11}$ & 0.099 & 9.1 & 0.19 \\
$[6.4,12.8] \times 10^{11}]$ & 230429 & $8.92 \times 10^{11}$ & 0.16 &$9.05\times 10^{11}$ & 0.16 & 8.3 & 0.33 \\
$[2.56,5.12]\times 10^{12}$ & 67400 & $3.56 \times 10^{12}$ &0.25& $3.62\times 10^{12}$ & 0.25 & 7.5 & 0.74 \\
$[1.02,2.048]\times10^{13}$ & 18322 & $1.42 \times 10^{13}$ & 0.39 & $1.45\times 10^{13}$ & 0.40 & 6.7 & 1.28 \\
$[4.1,8.192]\times 10^{13}$ & 4274 & $5.63 \times 10^{13}$ & 0.62 & $5.79\times 10^{13}$ & 0.63 & 5.7 & 2.20\\
$[1.64,3.277] \times 10^{14}$ & 650 & $2.22 \times 10^{14}$ & 0.98 & $2.32\times 10^{14}$ & 1.00 & 4.7 & 3.78 \\
\hline
\end{tabular}
\label{tab:halos}
\end{table*}

The shape of the halo-mass cross-correlation is clearly defined by two parts,
commonly referred to as the one-halo and two-halo terms since they are
dominated by particles within the same halo and in different halos,
respectively.  Figure~\ref{fig:xigm_mlink} shows that on large scales $\xihm$
follows closely the mass auto-correlation function, $\xi_{\rm mm}$, with a
mass-dependent offset in amplitude known as the halo bias factor, $b(M)$.  The
transition between the two regimes is remarkably sharp and takes place at an
overdensity of approximately seven times the mean density, i.e., $\xi_{\rm hm}
\simeq 6$.

We construct a simple model for $\xi_{\rm hm}(r)$ as follows:
\begin{eqnarray}
 \xi_{\rm model}(r; M) & =&
  \begin{cases}
\xioneh & \text{if $\xioneh \geq \xitwoh$}, \\
\xitwoh & \text{if $\xioneh < \xitwoh$},
  \end{cases} 
\label{eq:xihmmodel1}\\
\xioneh & = & \frac{\rho_{\rm halo}(r; M) - \rhomean}{\rhomean} 
\label{eq:xihmmodel2}\\
\xitwoh & = & b(M) \xi_{\rm lin}(r),
\label{eq:xihmmodel3}
\end{eqnarray}
where $\xi_{\rm lin}(r)$ is the mass autocorrelation function predicted by
linear theory.  The main ingredients of the model are the halo density profile,
$\rho_{\rm halo}$, and the bias factor $b(M)$ which we now describe in detail.

\subsection{The one-halo term}
\label{sec:onehalo}

The density profiles of CDM halos have been studied extensively with high
resolution N-body simulations over the past decade.  Early results indicated
that the density increases steeply towards the centres of halos \citep{FRENK85,
QUINN86, DUB91}.  \citet[][hereafter NFW]{NFW96,NFW97} suggested the following
simple fitting formula to describe the density profile of simulated halos:
\begin{equation}
\frac{\rho_{\rm NFW}(r)}{\rho_{\rm crit}}=\frac{\delta_0}{(r/r_s) (1+r/r_s)^{2}},
\label{eq:nfw}
\end{equation}
where $\rho_{\rm crit}$ is the critical density.  Note that the slope of the
NFW profile is shallower (steeper) than the isothermal profile inside (outside)
the characteristic scale radius, $r_s$.  Integrating this density
profile out to the virial radius, $r_{200}$, gives the following relation for
the dimensionless density parameter
\begin{equation}
\delta_0 = \frac{200}{3} \frac{c_{200}^3}{\ln(1+c_{200}) -
c_{200}/(1+c_{200})}
\label{eq:deltac}
\end{equation}
where the concentration parameter $c_{200} =r_{200}/r_s$.  Since the halo mass
and virial radius are related through $M_{200} = 200~\rho_{\rm crit}
(4\pi/3)~r_{200}^3$, the independent parameters in the NFW profile are
effectively the halo mass and concentration.  Furthermore, these properties
are known to be correlated in simulated halos, in the sense that low-mass
halos are more central concentrated than high-mass halos.  This is generally
interpreted in terms of the mean density of the universe at the time of
formation of a halo.  Since low-mass systems typically collapse at higher
redshift, the characteristic density and concentration of such systems is
larger with respect to high-mass systems.  The concentration-mass relation has
been studied extensively with cosmological simulations and several authors
have proposed models for predicting the average value of $c_{200}$ as a
function of halo mass and redshift \citep{NFW97,BULLOCK01,ENS01,MACCIO07,
NETO07,GAO07}.  Adopting such a model therefore fully specifies the density
profile of a typical halo of a given mass $M_{200}$ at a given redshift.

The most recent high resolution N-body simulations have revealed small but
significant deviations from the NFW formula.  \citet[][hereafter
N04]{NAVARRO04} show that the density profiles become shallower towards the
halo centre more gradually than the NFW formula predicts, causing NFW fits to
underestimate the density in the inner regions.  These authors propose an
improved fitting formula with the following form:
\begin{equation}
\ln \frac{\rho_{\alpha}}{\rho_{-2}} = \frac{-2}{\alpha} \left[\left(\frac{r}{r_{-2}}\right)^{\alpha} -1 \right],
\label{eq:rhoalpha}
\end{equation}
where $\rho_{-2}$ and $r_{-2}$ are the density and radius at which the
logarithmic slope of the density profile ${\rm d} \log \rho_{\alpha}/{\rm d}
\log r = -2$, and the parameter $\alpha$, controls the rate of change of the
logarithmic slope with radius.  Higher values of $\alpha$ cause the profile to
become shallower more quickly toward the centre of the halo.  Unlike the NFW
profile, eq.~\ref{eq:rhoalpha} does not converge to a power law at small
radii, instead reaching a finite density at the centre.  N04 found a
relatively small range of values, $0.12 \leq \alpha \leq 0.22$, to fit the
density profiles of simulated halos over a wide range in mass ($8 \times
10^9~\Msolh \lsim M_{200} \lsim 8 \times 10^{14}~\Msolh$).  A spatial density
profile of this form was first proposed in a different context by
\cite{EINASTO63} so we will hereafter refer to it as the Einasto profile.
\cite{PRADA06} recently used the NFW and Einasto formulae to fit halo density
profiles out to radii far beyond the virial radius.  These authors find 
the mean density to give a significant contribution to the density profile
at large distances, i.e., the halo profile is better fit by
\begin{equation}
\rho_{\rm halo}(r) = \rho_{\alpha}(r) + \rhomean. 
\end{equation}
With this modification, the one-halo term becomes
\begin{equation}
\xi_{\rm 1h}(r) = \frac{\rho_{\alpha}(r)}{\rhomean}.
\label{eq:xi1h}
\end{equation}
In subsections~\ref{sec:fits} and \ref{sec:densprof} we investigate the
accuracy of our model for $\xi_{\rm hm}$ using the NFW and Einasto fitting
formulae in the one-halo term of the model.

\subsection{The two-halo term}
\label{sec:twohalo}

The two-halo term of the model is specified by the bias factor, $b(M)$, and
the mass autocorrelation function calculated from linear theory, $\xi_{\rm
lin}(r)$.  Halo bias has been studied extensively in the context of
hierarchical structure formation scenarios.  Assuming a Gaussian distribution
of initial density fluctuations specified by $\sigma(M)$, the {\it rms} linear
mass fluctuation (extrapolated to $z=0$) within spheres that on average
contain mass $M$, \cite{MO96} derive an analytic model for halo bias, $b(\nu,
z)$, where
\begin{equation}
\nu = \left[ \frac{\delta_c(z)}{\sigma(M)} \right]
\label{eq:peakheight}
\end{equation}
is the dimensionless amplitude of fluctuations, or peak height, that produces
halos of mass $M$ at redshift $z$ and $\delta_c(z)$ is the linear overdensity
(again extrapolated to $z=0$) for which a spherical perturbation would
collapse at redshift $z$..  The characteristic halo mass for clustering
$M_*(z)$ is defined by $\sigma(M_*) = \delta_c(z)$ and halos more (less)
massive than $M_*$ are more (less) strongly clustered than the underlying mass
density field.  For the Millennium Simulation cosmology, $M_*(z=0) = 6.15
\times 10^{12}~\Msolh$.

\cite{MO96} showed that their model accurately describes the bias in the
autocorrelation function of dark matter halos with respect to the that of the
mass in cosmological N-body simulations.  Further testing against higher
resolution simulations has led to modifications and improvements in the model
\citep{JING98,GOVERNATO99, SHETH99,KRAVTSOV99, COLBERG00,SHETH01, SELJAK04,
MANDELBAUM05}.  Most recently, \cite{GAO05} compared these models with halo bias
in the Millennium Run.  Figure~1 of \cite{GAO05} shows that the Millennium Run
results are reasonably well-matched by the bias model of \cite{SHETH99}
\begin{equation}
b(\nu) = 1 + \frac{a\nu^2 - 1}{\delta_c} + \frac{2p}{\delta_c (1+(a\nu^2)^ p)}, 
\label{eq:bnu}
\end{equation}
with the parameter values $a=0.73$ and $p=0.15$ of \cite{MANDELBAUM05} based
on fits to the simulations of \cite{SELJAK04}.  We therefore adopt
eq.~\ref{eq:bnu} with these parameter values as the bias formula in the
two-halo term of our model for $\xi_{\rm hm}$.

\subsection{Model fitting results}
\label{sec:fits}

\begin{figure}
\plotone{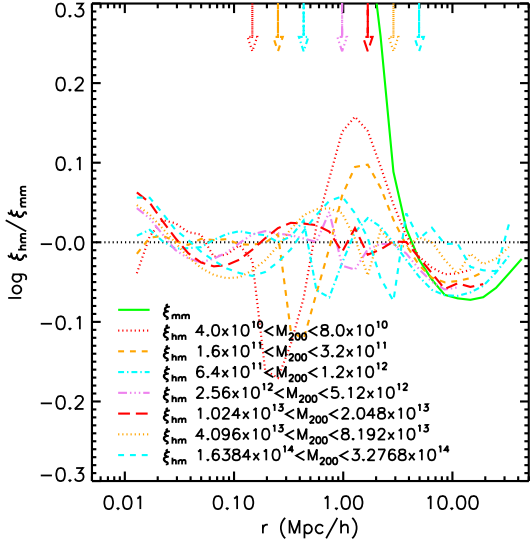}
\caption{Deviations between our measured halo-mass cross-correlations and the
simple model given by eqs.~\ref{eq:xihmmodel1}-\ref{eq:xihmmodel3} assuming
the NFW profile in the one-halo term.  On large scales, $r \gsim 3~\Mpch$, the
deviations are dominated by a quasi-linear distortion that is apparent in the
ratio of the mass autocorrelation function to the linear theory prediction,
$\xi_{\rm mm}/\xi_{\rm lin}$ (solid line).  For low mass halos, $M_{200} \lsim
3.2 \times 10^{11}$, the model also fails on intermediate scales suggesting a
cutoff in power should be applied to the linear prediction on these scales.
Otherwise the accuracy of the model better than $\simeq 10\%$.
\label{fig:xihmdev}}
\end{figure}

Figure~\ref{fig:xihmdev} shows the accuracy of our model when we choose the
halo density profile to be of NFW form.  The downward pointing arrows indicate
the transition scale $r_{\rm trans}$ between the one-halo and two-halo regimes
of the model, also listed in Table~\ref{tab:halos}.  We take the halo mass in
the model to be the geometric mean of the upper and lower halo mass limits in
each halo sample, i.e., $M_{\rm model} = (M_{\rm lo} M_{\rm hi})^{1/2}$.  Here
we take the concentration to be a free parameter and compare our best fit
values to the concentration-mass relations proposed by various authors.  The
best fit value is obtained by minimizing the root mean square of $(\xi_{\rm
hm} - \xi_{\rm model})/\xi_{\rm hm}$.

Several trends are apparent in the accuracy of our model fits.  In the
one-halo regime, the NFW profile fits the halo density profile to within about
$10\%$.  There is some indication of the systematic ``U-shaped'' residuals
found by N04 for fits to individual halo density profiles with the NFW
formula.  This indicates that the shape of the NFW profile does not perfectly
capture the simulation results.  We return to this issue later,
when we examine the accuracy of fits using eq.~\ref{eq:rhoalpha} to model the
halo density profile.

In the two-halo regime, the deviations in the model at large separations, $r
\gsim 3~\Mpch$, are dominated by the quasi-linear distortion due to the
large-scale movement of halos with respect to each other.  This is illustrated
by the solid line in Figure~\ref{fig:xihmdev} which shows the ratio of the
mass autocorrelation function to the linear theory prediction, $\xi_{\rm
mm}/\xi_{\rm lin}$.  This distortion has been investigated in numerous studies
of the matter power spectrum \citep{SELJAK00, MA00, SCOCCIMARRO01, SMITH03,
COLE05} and various correction factors have been proposed.  In this work we
prefer to neglect the distortion and adopt the simpler assumption of pure
linear theory \citep{PEACOCK00}.  This approach simplifies the calculation of
the model and avoids the introduction of the additional parameters that are
present in correction factors proposed by \cite{SMITH03} and \cite{COLE05} at
the cost of a systematic residual in our model fits on the order of $25\%$
at $\sim 10~h^{-1}$Mpc.

\begin{figure}
\plottwovert{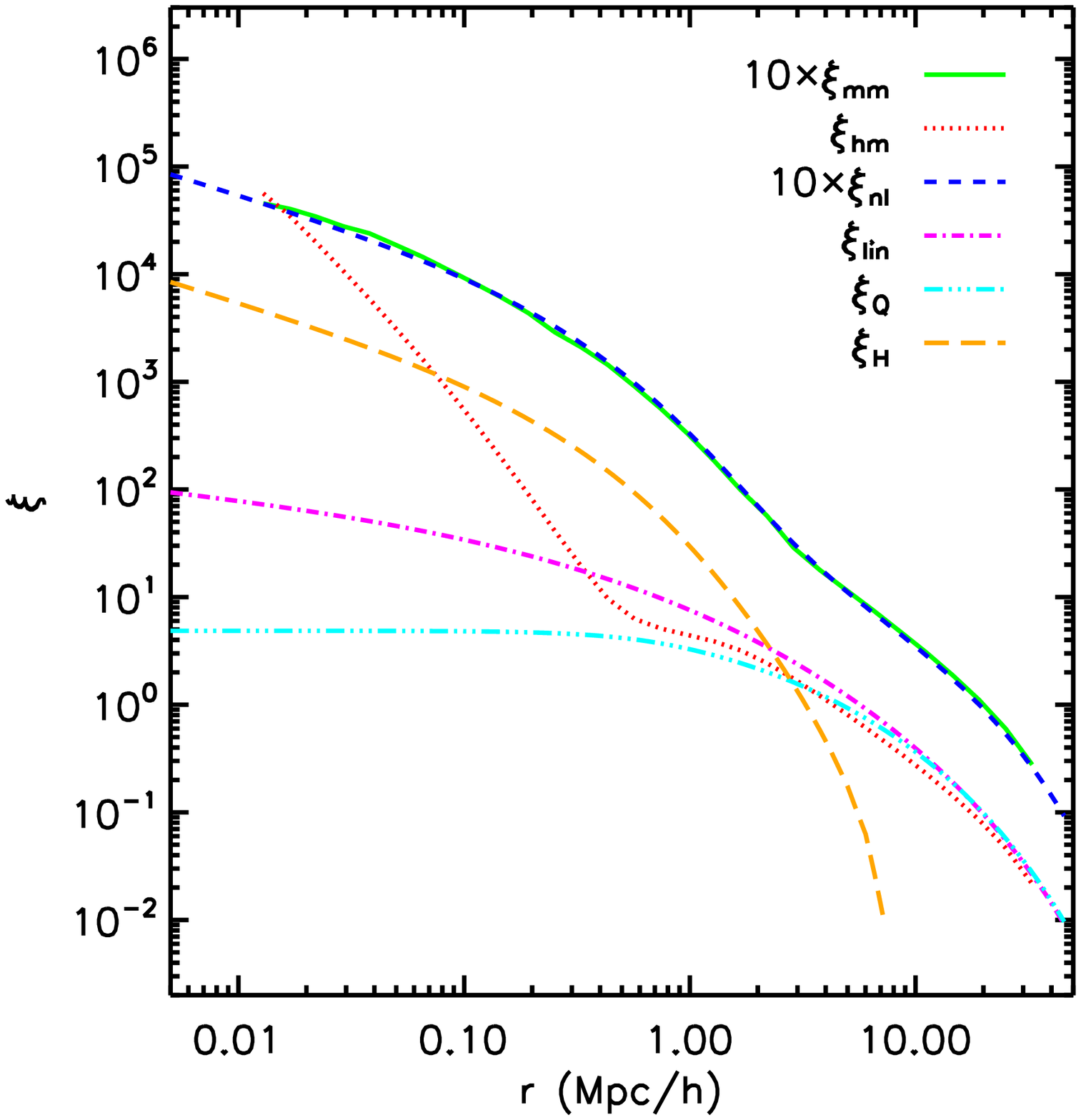}{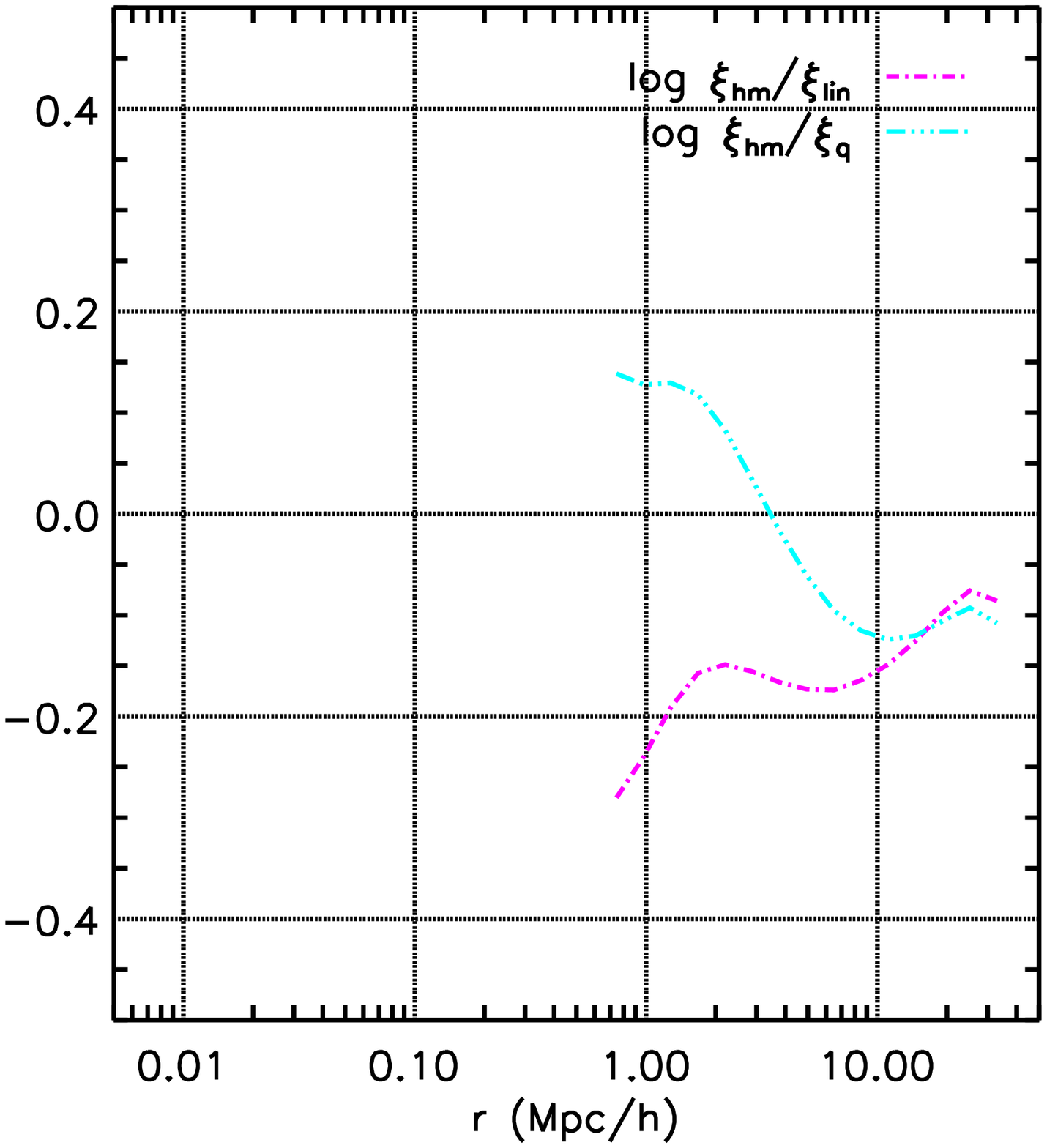}
\caption{ {\it Top panel:} Mass autocorrelation function, $\xi_{\rm mm}$, and
halo-mass cross-correlation functions, $\xi_{\rm hm}$, for low-mass halos
($4.0 \times 10^{10} < M_{200} < 8.0 \times 10^{10}$).  Also shown are the
predictions of linear theory $\xi_{\rm lin}$ and of the \citet{SMITH03}
HALOFIT model.  Neither $\xi_{\rm lin}$ nor the HALOFIT quasilinear model
$\xi_{\rm Q}$ provide a good match to the shape of $\xi_{\rm hm}$ on large
scales.  {\it Bottom panel:} The ratio between $\xi_{\rm hm}$ and the linear
and quasilinear predictions shows that the {\it shape} of $\xi_{\rm hm}$ is
marginally better fit by $\xi_{\rm lin}$ than by $\xi_{\rm Q}$.
\label{fig:halofit}}
\end{figure}

In the two lowest mass halo samples shown in Figure~\ref{fig:xihmdev}, the
two-halo term in the model appears to overpredict $\xi_{\rm hm}$ at
overdensities corresponding to $\xi_{\rm hm} \gsim 7$.  This suggests that the
linear theory prediction should also be modified by a cutoff in power at small
scales, as suggested by \cite{SMITH03}.  We investigate this in further detail
in Figure~\ref{fig:halofit}, where we examine the linear and quasi-linear
terms predicted by the \cite{SMITH03} HALOFIT model for the Millennium
Simulation cosmology.  The correlation function is related to the power
spectrum by the integral relation
\begin{equation}
\xi(r) = \int \Delta^2(k) \frac{\sin kr}{kr}\frac{dk}{k},
\end{equation}
and the HALOFIT model power spectrum is given by the sum of the quasi-linear
and one-halo terms,
\begin{equation}
\Delta^2(k) = \Delta_{\rm Q}^2(k) + \Delta_{\rm H}^2(k), 
\end{equation}
where the quasi-linear term $\Delta_{\rm Q}^2(k)$ includes an exponential
cutoff at the nonlinear wavenumber $k_\sigma \simeq 0.3~h~{\rm Mpc}^{-1}$ for
\LCDM.

The top panel of Figure~\ref{fig:halofit} shows that the correlation function
calculated from the HALOFIT model power spectrum provides a good fit to the
mass autocorrelation function $\xi_{\rm mm}$.  The cutoff in power at high $k$
built into the HALOFIT model corresponds to a flattening of the corresponding
correlation function, $\xi_{\rm Q}$ with respect to the linear theory
prediction, $\xi_{\rm lin}$ at an overdensity $\xi_{\rm Q}\simeq 4$.
\cite{SMITH03} provide fitting formulae for $\Delta_{\rm Q}^2(k)$ and
$\Delta_{\rm H}^2(k)$, each with numerous parameters tuned to give a good
match to the matter power spectrum in various CDM-based cosmological N-body
simulations.  However, Figure~\ref{fig:halofit} shows that $\xi_{\rm Q}$
cannot be used to improve the model for $\xi_{\rm hm}$ of low-mass halos.
This is illustrated in the bottom panel, where we show the ratios $\xi_{\rm
hm}/\xi_{\rm Q}$ and $\xi_{\rm hm}/\xi_{\rm lin}$ for halos with mass $4.0
\times 10^{10}~\Msolh < M_{200} < 8.0 \times 10^{10}~\Msolh$.  We find that
using the quasi-linear term from the HALOFIT model actually produces a
marginally poorer fit to the measured shape of halo-mass cross-correlations.
The HALOFIT model imposes a cutoff in power at the non-linear wavenumber
$k_\sigma$. Our results suggest that the \cite{SMITH03} prescription adopts a
value of $k_\sigma$ that is too low in the present context.  Modifying their
prescription for the non-linear scale would require the recalibration of all
of the fitting formulae in the HALOFIT model in order to recover the fit to
the mass autocorrelation function. This is beyond the scope of the present
paper.

\subsection{Halo concentrations}
\label{sec:haloconc}

\begin{figure}
\plotone{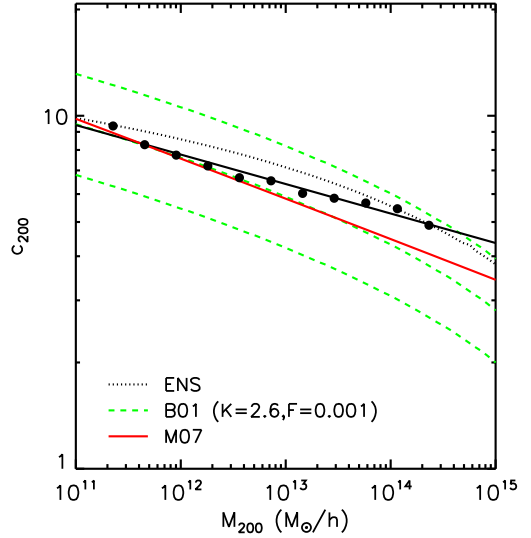}
\caption{NFW concentration versus halo mass from $\xi_{\rm hm}$ model fits
compared with the predictions of various models.  The solid line shows the
best-fit power law with $c \propto M^{-0.080}$. }
\label{fig:cvirplt}
\end{figure}

Figure~\ref{fig:cvirplt} shows the best-fit NFW concentration parameter values
versus halo mass for our model fits.  Also shown are the predictions of the
concentration-mass models proposed by \citet[][hereafter ENS]{ENS01},
\citet[][hereafter B01]{BULLOCK01}, and the power-law fits of
\citet[][hereafter M07]{MACCIO07} and \citet{NETO07}.  Our best-fit values
appear to follow the B01 and M07 results for halo masses $M_{\rm vir} \lsim
5\times10^{12}$, but at higher masses the results are better described by the
ENS model.  We find that a power-law provides a reasonable fit to the data,
with the same normalization, but a slightly shallower slope ($c \propto
M^{-0.080}$) than found by M07 and \cite{NETO07} ($c \propto M^{-0.109}$). We
note that the difference between all these models is relatively small with
respect to the $1-\sigma$ scatter, $\Delta \log (c_{\rm vir}) = 0.14$, found
for individual halos in B01.  Since the ENS model provides a reasonably close
match to our best-fit concentration values and has been calibrated for
redshifts $z > 0$ and for different cosmologies, we adopt it as our halo
concentration model hereafter.

\subsection{Halo density profile}
\label{sec:densprof}

In this subsection we investigate whether our model is improved if we change
the choice of halo density profile.  We here use the Einasto profile
(eq.~\ref{eq:rhoalpha}), the improved fitting formula proposed by N04.  This
contains an additional shape parameter, $\alpha$, which controls the rate at
which the slope of the density profile changes with radius, with typical
values lying in the range $[0.12, 0.22]$ \citep[N04,][]{PRADA06}.

\begin{figure}
\plotone{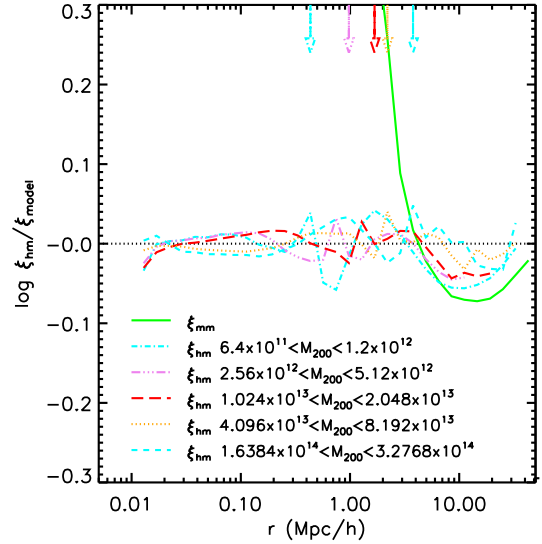}
\caption{Deviations between our measured halo-mass cross-correlations and the
simple model given by eqs.~\ref{eq:xihmmodel1}-\ref{eq:xihmmodel3} with the
Einasto profile in the one-halo term.  On small scales the Einasto profile
provides a better fit than the NFW profile, with deviations $\lsim 5\%$.
\label{fig:xihmdevalpha}}
\end{figure}

\begin{figure}
\plotone{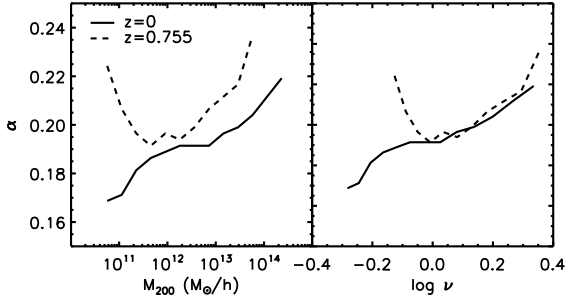}
\caption{Best-fit value of $\alpha$ from the Einasto fitting formula versus
halo mass.  Higher values of $\alpha$ correspond to density profiles that
becomes shallower more quickly towards halo centre.  The best-fit value of
$\alpha$ tends to increase with halo mass.}
\label{fig:alphamvir}
\end{figure}

Figure~\ref{fig:xihmdevalpha} shows the deviations of the measured halo-mass
cross-correlations from our best fits using the Einasto formula.  Here we
focus on halo samples with $M_{200} > 6.4 \times 10^{11}~\Msolh$, since
deviations from the fits are then not dominated by the inadequacy of the
two-halo term on intermediate scales (see Section~\ref{sec:fits}).  The
deviations are reduced significantly by using the Einasto model, $\lsim 5\%$
compared to $\lsim 10\%$ for fits using the NFW profile.  The best-fit values
of $\alpha$ are shown as a function of halo mass in
Figure~\ref{fig:alphamvir}.  We find that $\alpha$ tends to increase with halo
mass, ranging from $\alpha \simeq 0.12$ for $M_{200} \simeq 10^{12}~\Msolh$ to
$\alpha \simeq 0.2$ for $M_{200} \simeq 3\times10^{14}~\Msolh$.

\subsection{Higher redshift results}
\label{sec:redshift}

Since galaxy-galaxy lensing studies are sensitive to lenses at redshifts well
above zero, it is interesting to check our models against cross-correlations
at $z>0$.  In the COSMOS survey, for example, the galaxy-galaxy lensing signal
is significant for lenses over the redshift range $0.2 < z < 1.2$ with the
main contribution from $z \simeq 0.5-0.9$ (A.~Leauthaud, private
communication).

\begin{figure}
\plottwovert{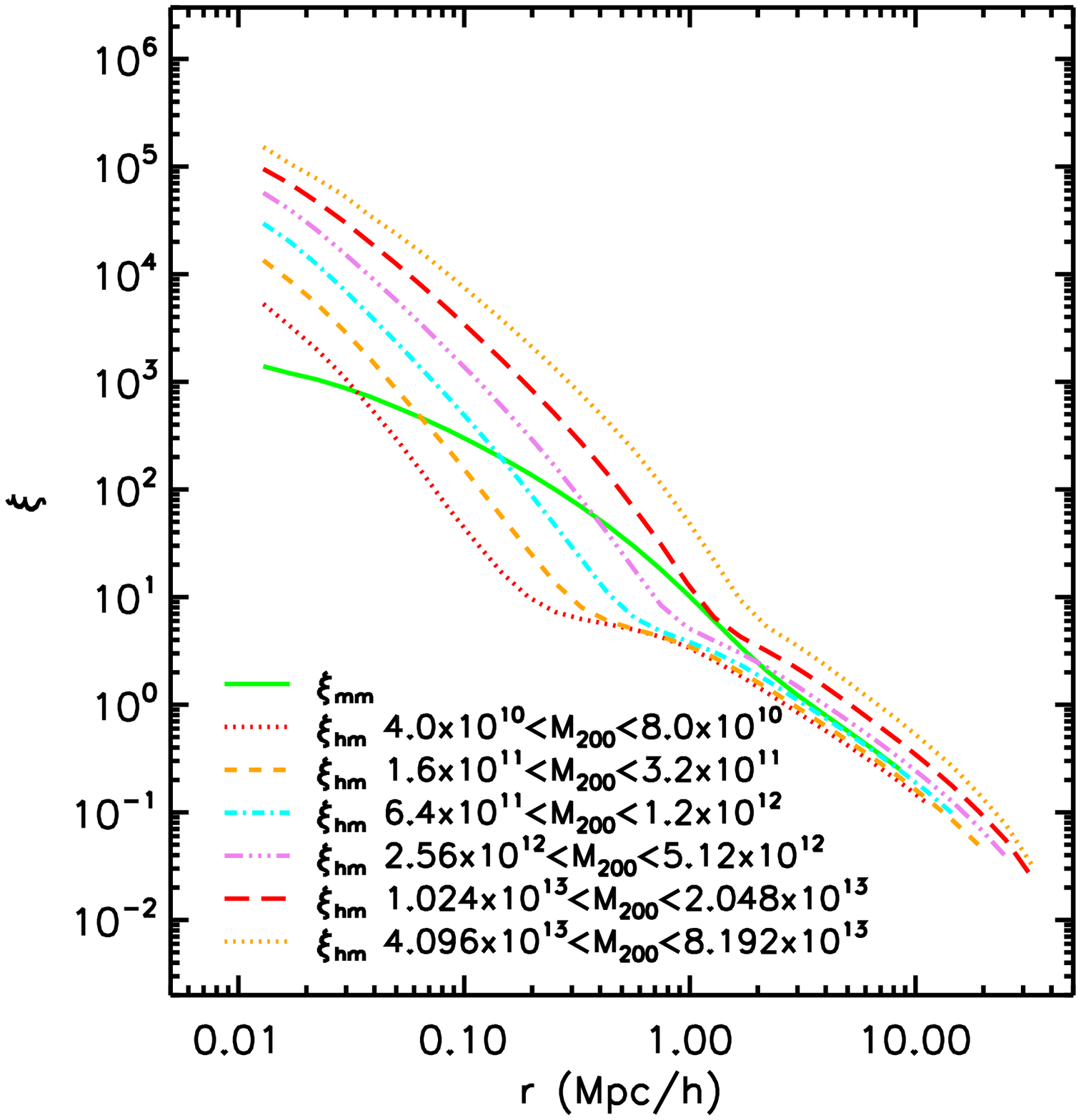}{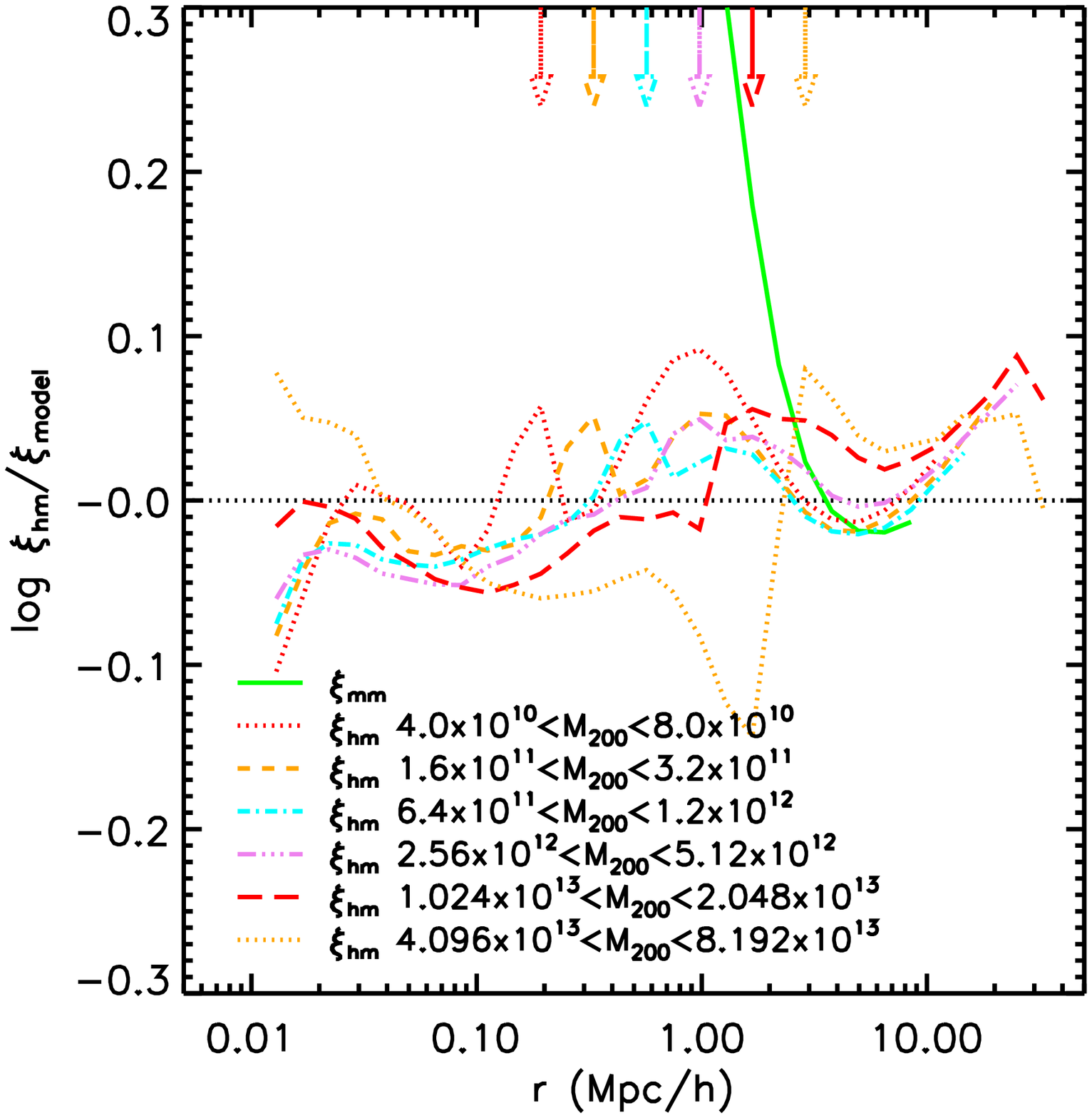}
\caption{{\it Top panel:} Halo-mass cross-correlations calculated at
$z=0.755$. {\it Bottom panel:} Deviations between $\xi_{\rm hm}$ and our model
assuming the ENS halo concentration relation and the Einasto density
profile. Deviations from the model are on the order of $10-20\%$. }
\label{fig:xihm44}
\end{figure}

Figure~\ref{fig:xihm44} shows $\xi_{\rm hm}$ for the $z=0.755$ output of the
Millennium Simulation.  The results are qualitatively similar to those at
$z=0$, and the bottom panel of the figure shows the deviations from the best
fits using our (Einasto) model.  We find that the deviations are on the order
$\pm 10\%$ for the higher redshift $\xi_{\rm hm}$.  Figure~\ref{fig:alphamvir}
also shows the best fit values of $\alpha$ for this redshift.  As in the $z=0$
case, we find that $\alpha$ tends to increase with mass for $M_{200} \gsim 3
\times 10^{11}~\Msolh$.  In the left panel of this figure we plot the best fit
$\alpha$ values against the peak height defined by eq.~\ref{eq:peakheight}.
Plotted in this way, we find that the $z=0$ and $z=0.755$ results are in good
agreement with each other.  We note that a similar result has been obtained by
Gao et al (in preparation) in a study of cluster halos from the Millennium
Simulation. This provides a simple way to estimate appropriate $\alpha$
values for other redshifts and masses.

\subsection{Shear}
\label{sec:shear}

Galaxy-galaxy lensing studies measure the average tangential distortion or
shear of background source galaxies due to foreground lenses.  If the
redshift distributions of the sources and lenses are known, the average
tangential shear, $\gamma_t$, is related to the surface mass density as
follows
\begin{equation}
\Sigma_{\rm crit} \gamma_t = \overline{\Sigma}(<R) - \Sigma(R) \equiv \Delta
\Sigma(R), 
\end{equation}
where $\overline{\Sigma}(<R)$ is the mean surface density within the projected
radius $R$, and the critical surface density is given by
\begin{equation}
\Sigma_{\rm crit} = \frac{c^2}{4 \pi G} \frac{D_s} {D_l D_{ls}}, 
\end{equation}
where $D_s$ and $D_l$ are the angular diameter distances to the lens
and the source, respectively, and $D_{ls}$ is the angular diameter distance
between the lens and the source.  In order to compare our results with
observational measurements, we calculate the surface mass density by projecting
the three-dimensional mass density:
\begin{equation}
\Sigma(R) = 2 \int_r^\infty \frac{\Delta \rho(r)}{\sqrt{r^2-R^2}} r {\rm d}r,
\end{equation}
where $\Delta\rho(r) \equiv \rho(r) - \overline{\rho}$. 

\begin{figure}
\plottwovert{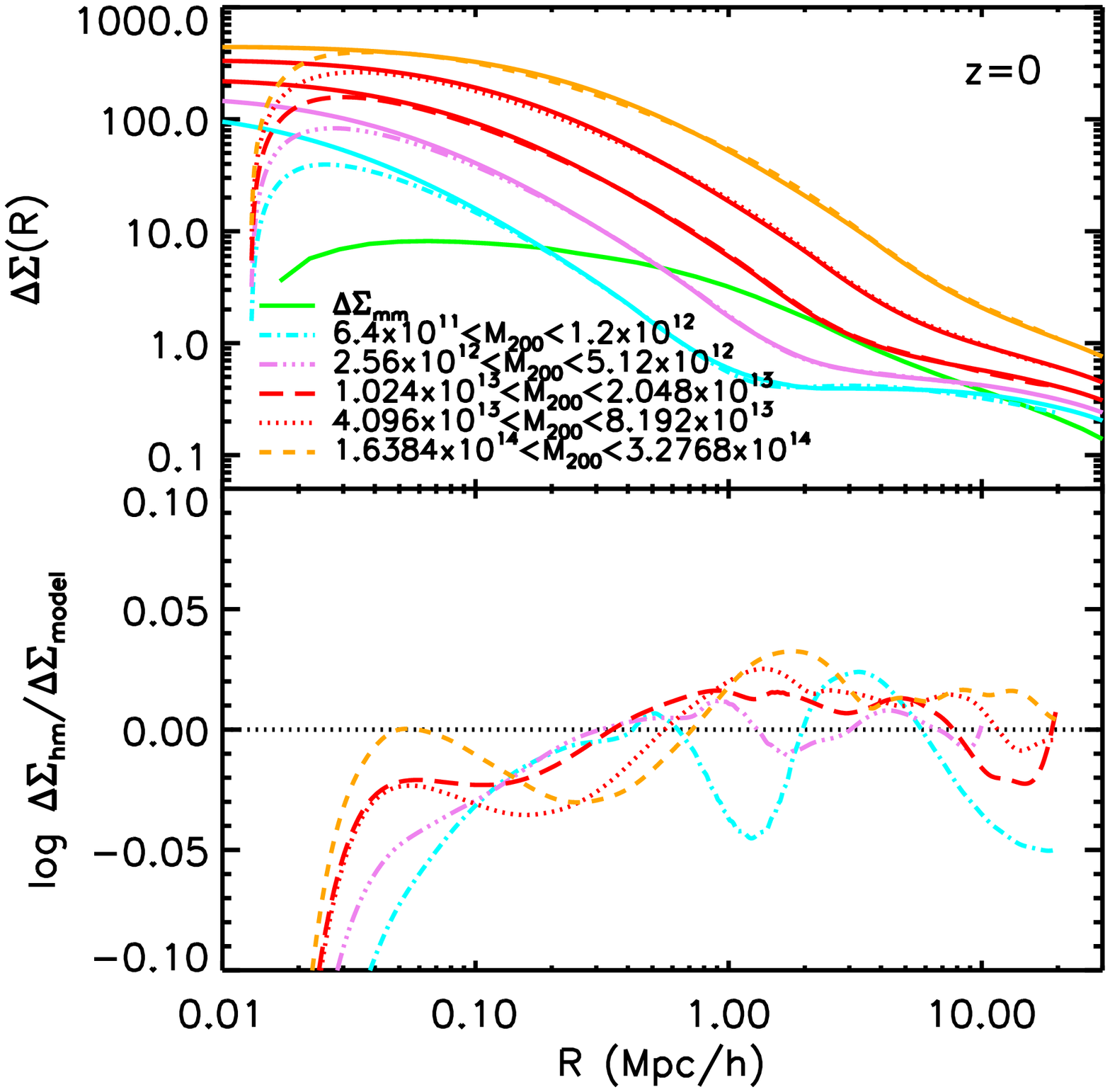}{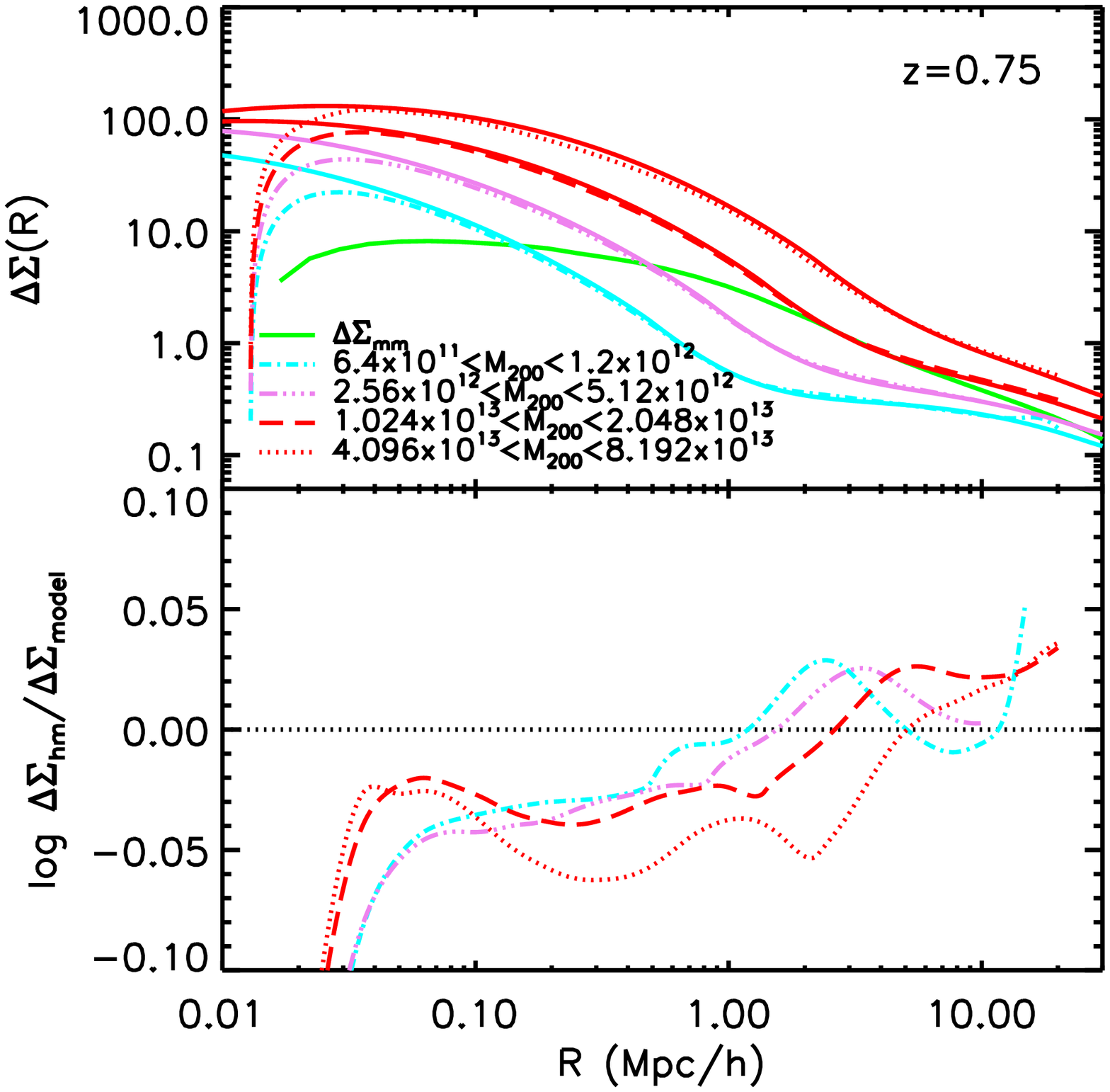}
\caption{{\it Top panels:} $\Delta \Sigma(R)$ calculated from $\xi_{\rm hm}$ at
$z=0$ and $z=0.755$.  Solid lines show $\Delta \Sigma(R)$ corresponding to best
fits to $\xi_{\rm hm}$.  {\it Bottom panels:} Deviations from our fits. The
downturn at small scales $r \lsim 0.03~\Mpch$ is due to integrating over scales
below the resolution limit of the simulation.  On scales larger than this, the
magnitude of the deviations are on the order of $10\%$ in terms of $\Delta
\Sigma(R)$.  }
\label{fig:sigmaplot}
\end{figure}

Figure~\ref{fig:sigmaplot} shows the quantity $\Delta \Sigma(R)$ calculated
from our $\xi_{\rm hm}$ curves and our model fits.  Note that the model fits
have not been recalculated to minimize the deviations in $\Delta \Sigma(R)$.
We find that down to a scale of $\simeq 0.03-0.05~\Mpch$, below which our
calculation of $\overline{\Sigma}(R)$ from the simulation becomes
unreliable, the deviations from the model fits are $\lsim 10\%$ for both $z=0$
and $z=0.755$.

\section{Galaxy-mass cross-correlations}
\label{sec:xigm}

We now investigate cross-correlations between galaxies and mass in the
Millennium Simulation.  As described in Section~\ref{sec:sim}, semi-analytic
models have been grafted onto the simulation in order to predict the evolution
of the galaxy population.  A central galaxy is associated with the main
subhalo of each FoF halo and satellite galaxies are associated with the other
subhalos.  In addition, there are so-called `orphan' galaxies.  These are
satellite galaxies whose associated subhalos have been tidally stripped to
below the resolution limit of the simulation.  Such a galaxy remains
identified with the individual particle which was the most bound member of the
subhalo at the last time it could be identified.  Orphan galaxies are assumed
to survive for an estimated dynamical friction time, after which they merge
with the central galaxy of their halo. They are a significant fraction of the
faint galaxies in the model. \cite{GAO04}, \cite{WANG06} and \cite{SALES07}
show that they must be included if the model is to predict the small-scale
clustering of galaxies accurately.

\subsection{Cross-correlations between central galaxies and mass}
\label{sec:centgal}

\begin{figure}
\plottwovert{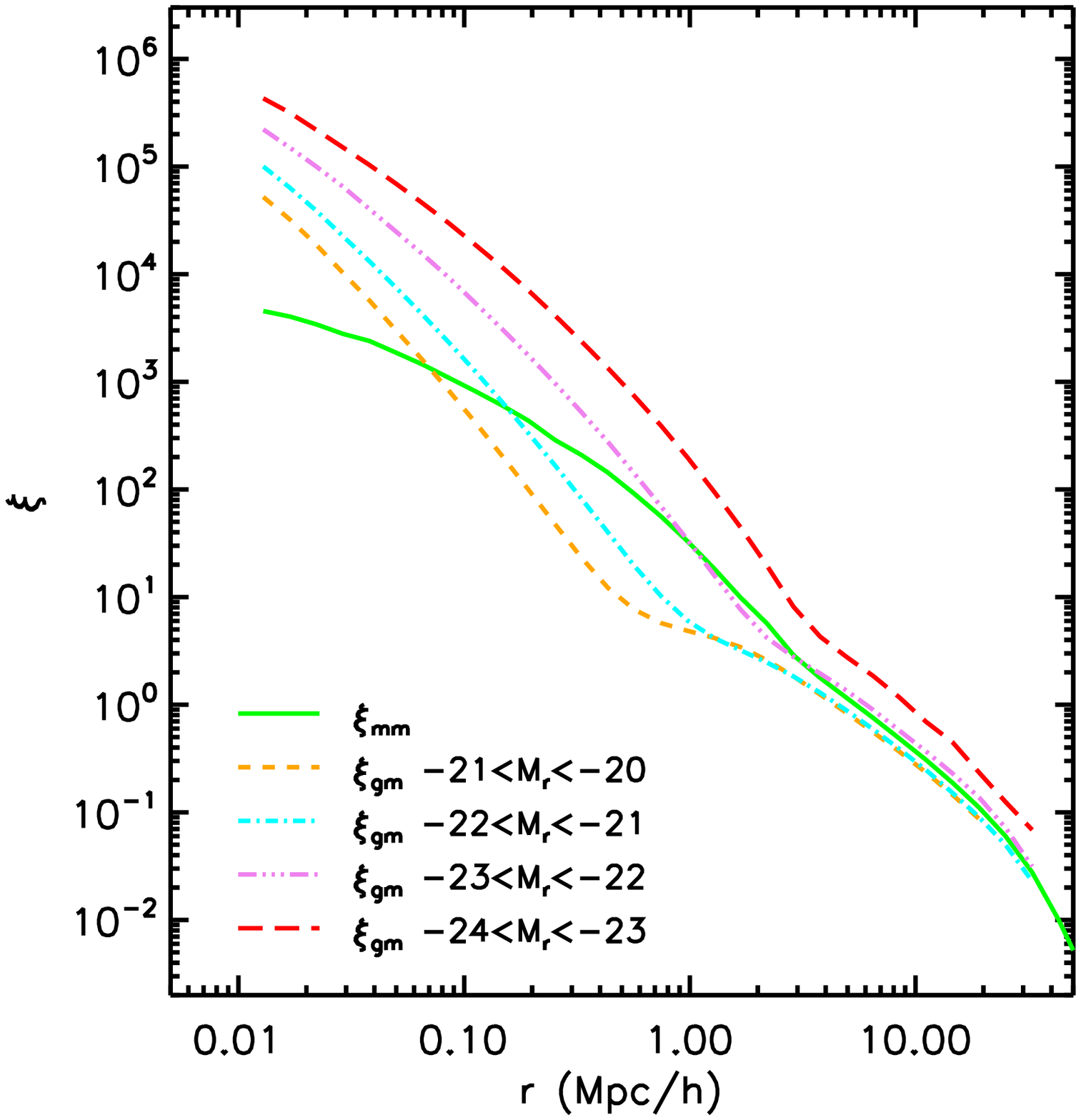}{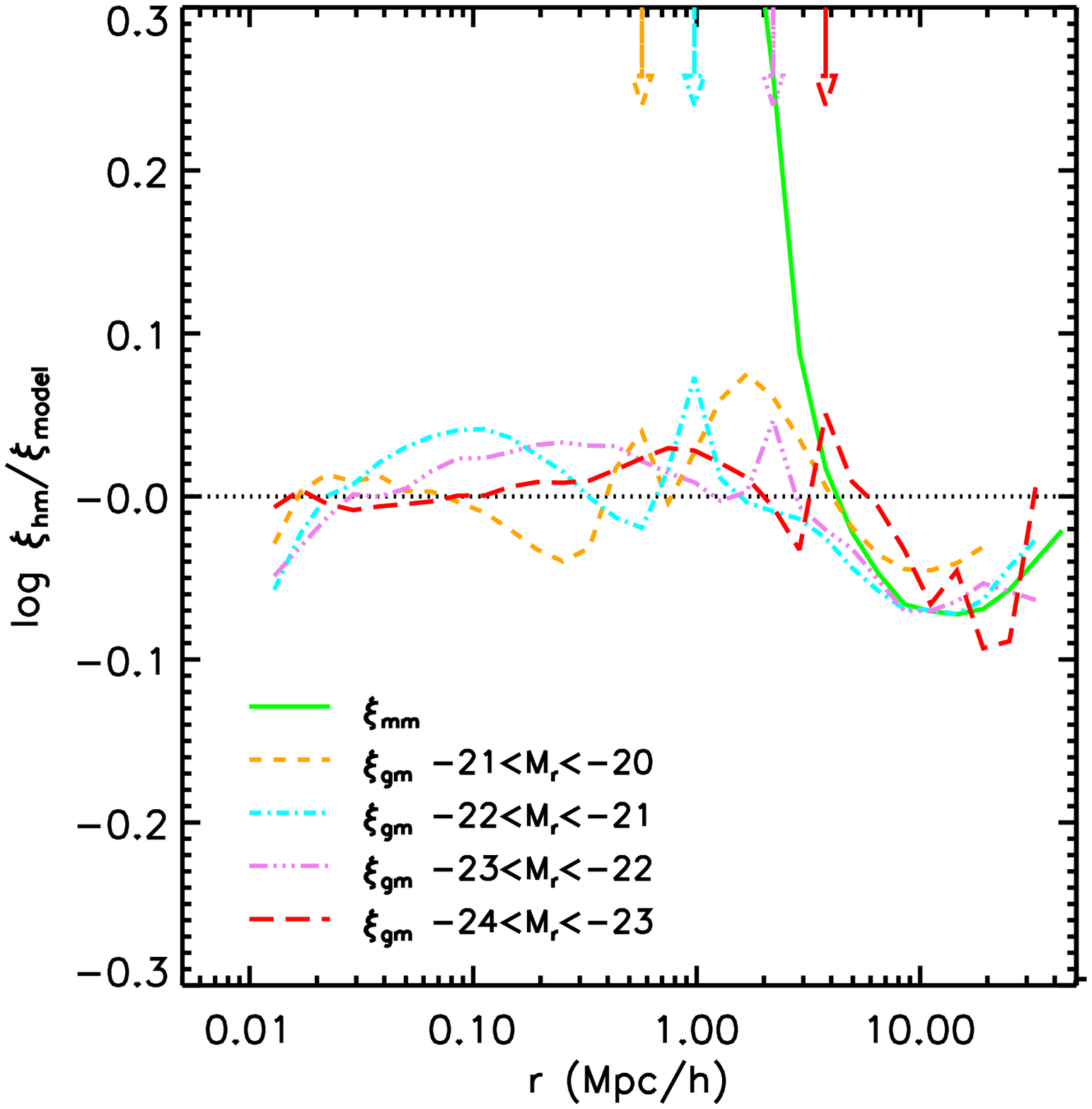}
\caption{ {\it Top panel:} Galaxy-mass cross-correlations at $z=0$ for central
galaxy samples selected according to r-band absolute magnitude.  {\it Bottom
panel:} Deviations from our model for halo-mass cross-correlations, where the
mean host halo mass has been taken as a free parameter.
\label{fig:xigm}}
\end{figure}

We first consider cross-correlations between central galaxies and mass.  Since
central galaxies are placed on the most bound particle of their host halos,
$\xi_{\rm gm}$ for central galaxies is equivalent to $\xi_{\rm hm}$ averaged
over the corresponding sample of FoF halos, and so should be similar to the
$\xi_{\rm hm}$ presented in Section~\ref{sec:xihm} provided that central
galaxies are selected according to a property that correlates well with host
halo mass.  Figure~\ref{fig:xigm} shows $\xi_{\rm gm}$ for central galaxies
selected according to r-band absolute magnitude in the range $-20 < M_r <
-24$.  These cross-correlations indeed appear very similar in shape to those
of Figure~\ref{fig:xigm_mlink}.

We apply the simple model presented in Section~\ref{sec:xihm} based on the
Einasto profile of eq.~\ref{eq:rhoalpha}.  We take the mass of the host halo
as a free parameter which determines the bias factor according to
eq.~\ref{eq:bnu} and the radius and concentration of the mean halo density
profile according to the ENS model.  The best fit value of the halo density
profile parameter $\alpha \simeq 0.15$ for the central galaxy $\xi_{\rm gm}$.
This is quite similar to the values found above for halos in the relevant
mass ranges.

\begin{figure}
\plottwovert{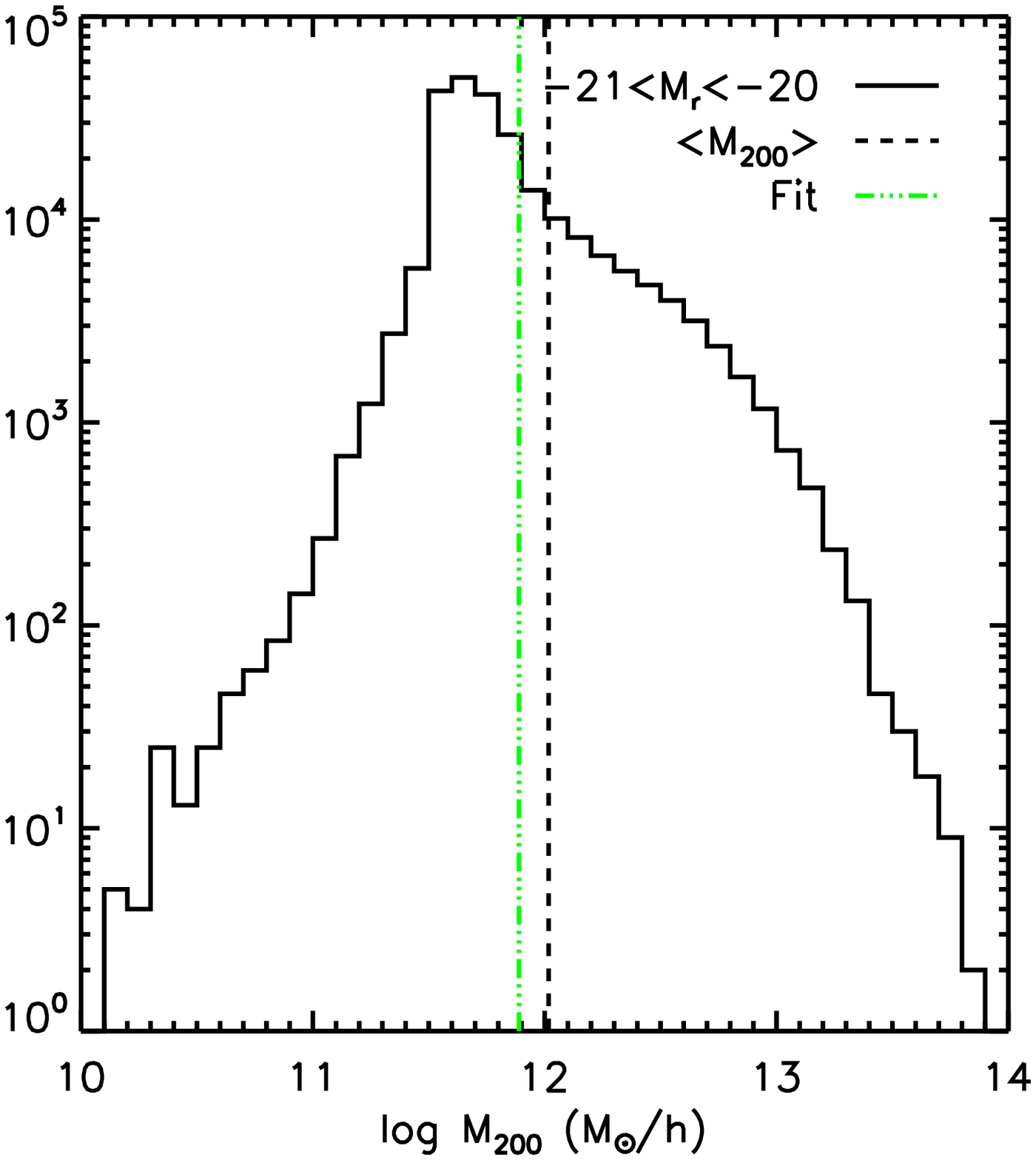}{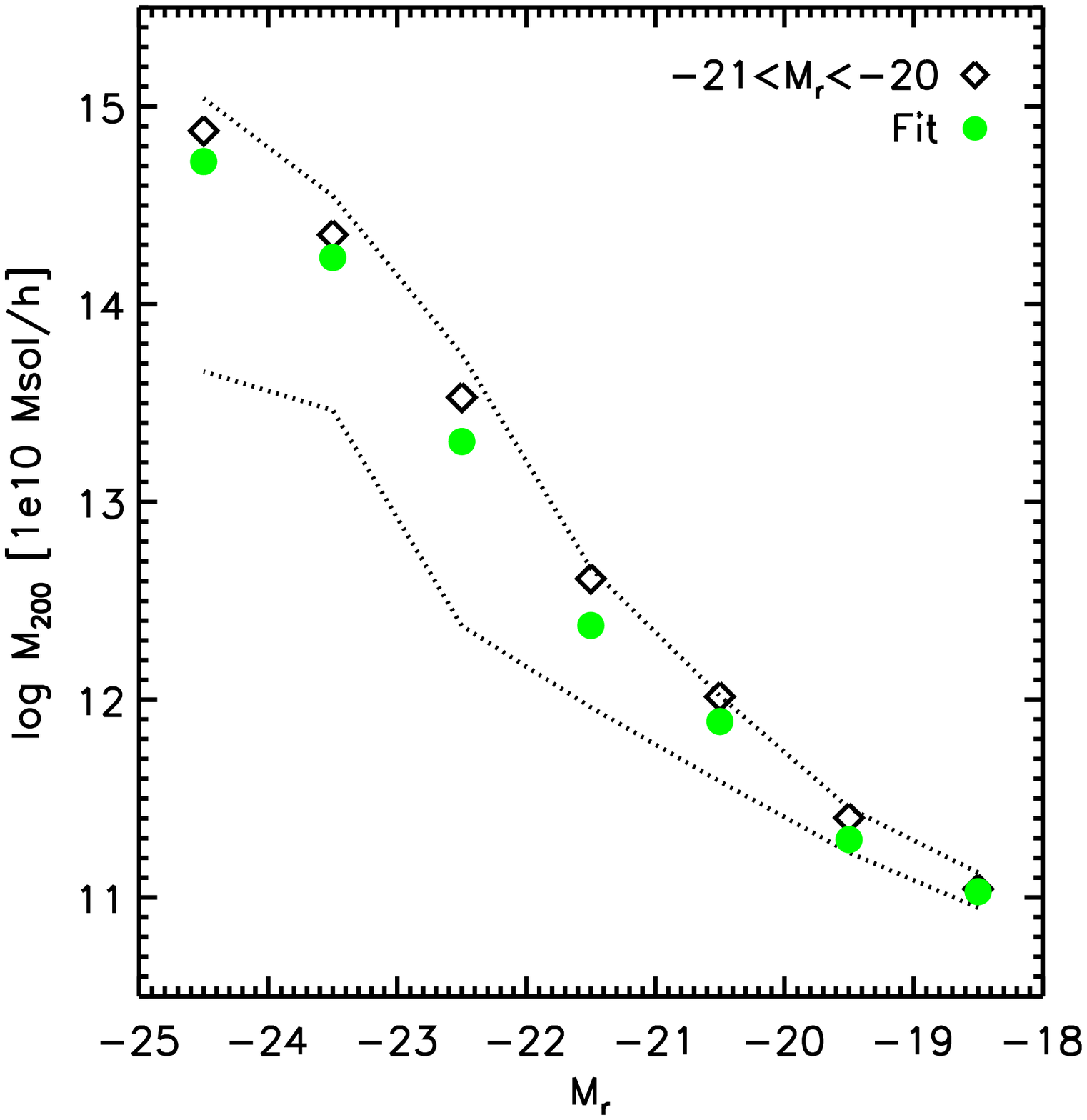}
\caption{ {\it Top panel:} Distribution of host halo mass $M_{200}$ for
central galaxies with $-21 < M_r < -20$.  {\it Bottom panel:} Mean host halo
$M_{200}$ (open diamonds) with 20-and 80-percentile values (dotted lines) are
compared with the best-fit values recovered by modelling galaxy-mass
cross-correlations (solid circles).
\label{fig:m200dist}}
\end{figure}

The distribution of host halo masses for central galaxies in the luminosity
range $-21 < M_r < -20$ is shown in Figure~\ref{fig:m200dist}.  This
distribution is highly asymmetric and ranges over four orders of magnitude
in mass. Nevertheless, the best fit model recovers the mean halo mass within 
about 30\%.  The bottom panel of the figure shows mean and best-fit halo
mass values as a function of $M_r$.  The fit values tend to underestimate the
mean halo mass, typically by about $30\%$.

\subsection{Cross-correlations between satellite galaxies and mass}
\label{sec:satgal}

We now consider cross-correlations between satellite galaxies and the mass.
These galaxies were once associated with the dominant subhalo of a FoF group
but are now centred on a secondary subhalo within a larger FoF group and may
have experienced significant mass loss due to tidal stripping.  Indeed, as
noted above, a significant fraction are `orphans' and have lost their
subhalo entirely, remaining associated with the particle which lay at subhalo
centre when it was last identified.  Observation shows that the baryonic
components of galaxies are substantially overdense with respect to the dark
matter, and simulations suggest that they are therefore more resistant to
tidal disruption \citep{KWH92,KWH96}.  The semi-analytic galaxy formation
model takes this into account by allowing orphans to survive for a dynamical
friction time before merging them with the central galaxies of their halos.

Table~\ref{tab:gals} shows the numbers of central and satellite galaxies in
the Millennium Simulation model of \cite{CROTON06} as a function of $M_{r}$.
Although there are fewer satellites than central galaxies for all $M_r \leq
-17$, the fraction of satellites increases with decreasing luminosity, and is
about $40\%$ of the galaxy population for $-19 < M_r < -17$.  The fraction of
orphans also increases with decreasing luminosity, and is more than half of
the total satellite population in the $-18 < M_r < -17$ magnitude range.

\begin{table*}
\centering
\caption{Galaxy properties.}
\begin{tabular}{lllllllll}
\hline
$M_r$ & $N_{\rm central}$ & $<M_{200}>_{\rm central}$ &
$N_{\rm subhalo}$ & $N_{\rm orphan}$ & $f_{\rm sat}$ & $<M_{200}>_{\rm host}$ &
$r_{\rm subhalo}$ & $r_{\rm orphan}$\\
& & $[\Msolh]$ & & & & $[\Msolh]$ & $[\Mpch]$ & $[\Mpch]$ 
\\ \hline
$[-17,-18]$ & 1927798 &$4.5\times 10^{10} $ &602965 & 752317 & 0.41 &$6.6\times 10^{13}$ &0.52 &   0.36\\
$[-18,-19]$ & 1252316 &$1.1\times 10^{11}$ &473054 & 404325 & 0.41  &$6.9\times 10^{13}$ &0.55 &   0.32\\
$[-19,-20]$ & 923701  &$2.0\times 10^{11}$ &308907 & 209794  & 0.35 &$7.7\times 10^{13}$ &0.59 &   0.29\\
$[-20,-21]$ & 600710  &$7.8\times 10^{11}$ &165187  & 67691  & 0.27 &$8.5\times 10^{13}$ &0.66 &   0.26\\
$[-21,-22]$ & 236524  &$2.4\times 10^{12}$ &35953  & 7988   & 0.15  &$1.0\times 10^{14}$ &0.83 &   0.23\\
$[-22,-23]$ & 15984   &$2.0\times 10^{13}$ &1553 & 372  & 0.10      &$2.3\times 10^{14}$ &1.35 &   0.17\\
$[-23,-24]$ & 617     &$1.7\times 10^{14}$ & 27  & 3 & 0.04         &$4.7\times 10^{14}$ &2.42 &   0.33\\
$[-24,-25]$ & 12      &$5.3\times 10^{14}$ &0  & 0  & 0             &$0$  & 0&0\\
\hline
\end{tabular}
\label{tab:gals}
\end{table*}

\begin{figure}
\plottwovert{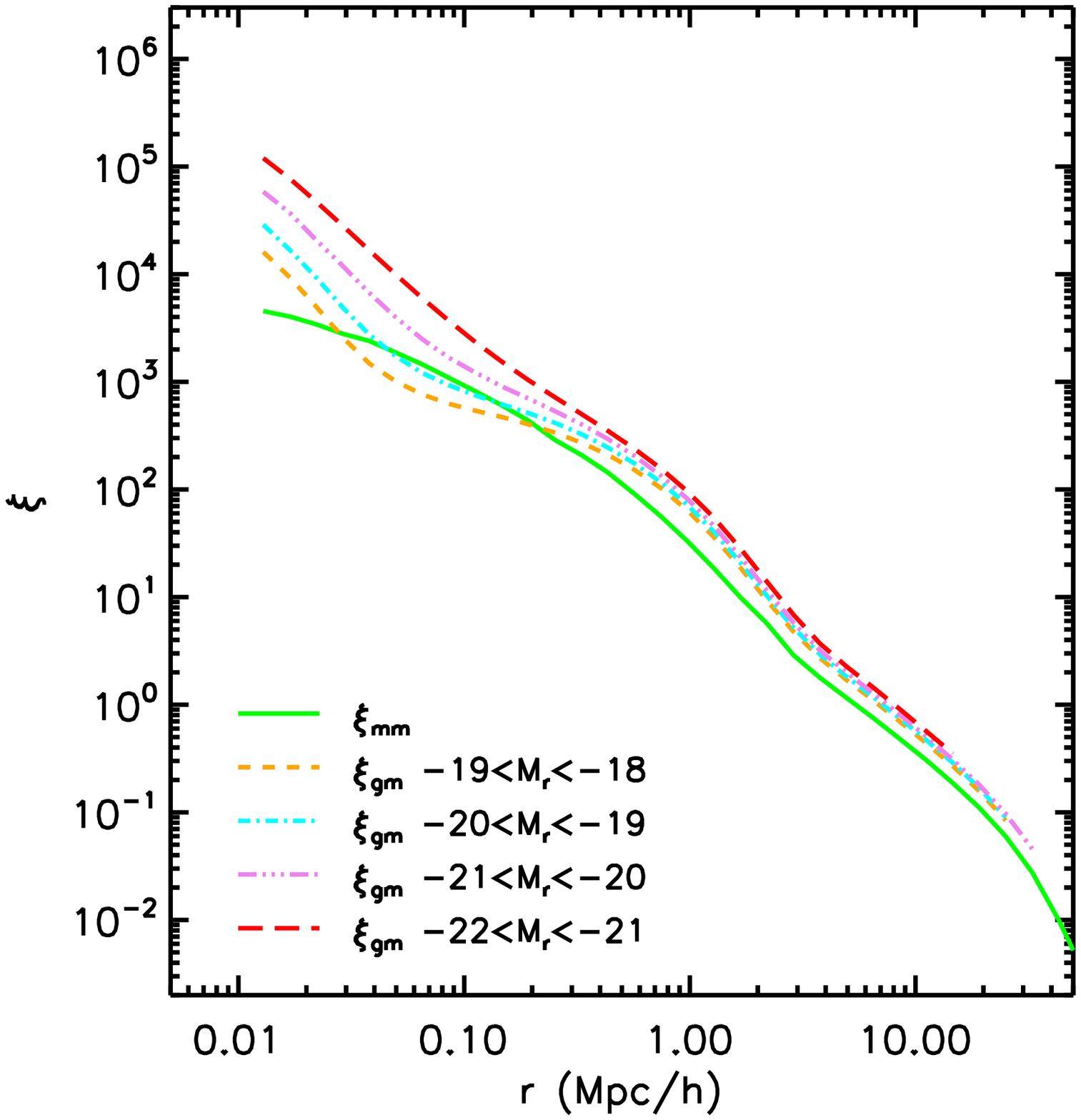}{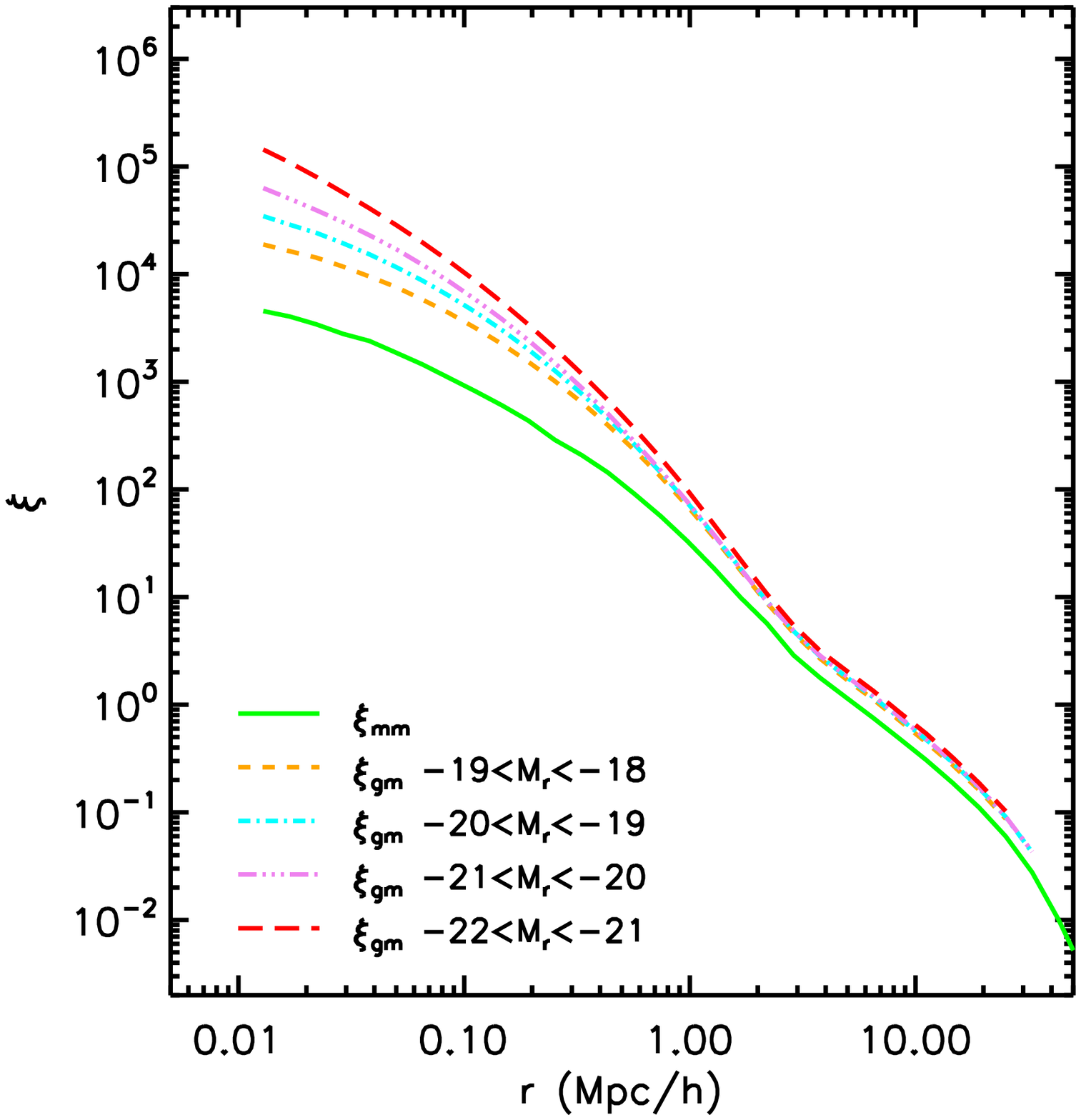}
\caption{Cross-correlation between the mass and satellite galaxies with ({\it
top panel}) and without ({\it bottom panel}) associated subhalos.  In both
cases, $\xi_{\rm gm}$ is positively biased on large scales, $r \gsim 1~\Mpch$
indicating that satellites preferentially reside in high mass halos.  On small
scales $\xi_{\rm gm}$ for subhalo satellites shows an upturn due to
the mass bound to the individual subhalos, whereas the orphan satellite
$\xi_{\rm gm}$ reflects the density profile of the host halo convolved with
the radial distribution of the orphan satellites.
\label{fig:xigmtype}}
\end{figure}

Figure~\ref{fig:xigmtype} shows $\xi_{\rm gm}$ for ``subhalo satellites'',
i.e., satellite galaxies hosted by intact subhalos (top panel) and for orphan
satellites (bottom panel).  For both types of satellite, the cross-correlation
function is positively biased with respect to $\xi_{\rm mm}$ on large scales,
$r \gsim 1~\Mpch$.  This reflects the fact that relatively bright satellite
galaxies almost all reside in halos more massive than $M_{*}$, the
characteristic mass at which halos are as strongly clustered as the underlying
mass density field.

On small scales, $r \lsim 0.1~\Mpch$, the shape of $\xi_{\rm gm}$ for
satellite subhalos reflects the mass associated with the individual subhalos.
A comparison between the $\xi_{\rm gm}$ curves in the top panel of
Figure~\ref{fig:xigmtype} and those in Figure~\ref{fig:xigm} shows that the
average density profile of a satellite galaxy with a given luminosity is quite
similar in both shape and amplitude to that of a central galaxy with the
same luminosity.  In detail, the subhalo satellite $\xi_{\rm gm}$ is
denser by $\simeq 10-20\%$ with respect to the corresponding central galaxy
$\xi_{\rm gm}$, reflecting the higher concentration of subhalos compared to
main halos \citep{BULLOCK01}.

In the case of orphan satellites, negligible mass is associated with the
satellite itself. On small scales the shape of the cross-correlation function
reflects that of the host halos after convolution with the mean radial
distribution of orphans within their hosts.  Since orphans are created by
tidal stripping, they are found predominantly close to the centres of their
hosts where tidal forces are strongest.  Their radial profile is therefore
highly centrally concentrated \citep{GAO04}. Table~\ref{tab:gals} shows that
the mean distance of orphan satellites from the centres of their hosts is
significantly smaller than that of subhalo satellites.  As a result, the
orphan satellite $\xi_{\rm gm}$ is overdense on small scales with respect to
$\xi_{\rm mm}$, since the one-halo term of the latter reflects the halo
density profile convolved with itself \citep{MA00}.

\begin{figure}
\plottwovert{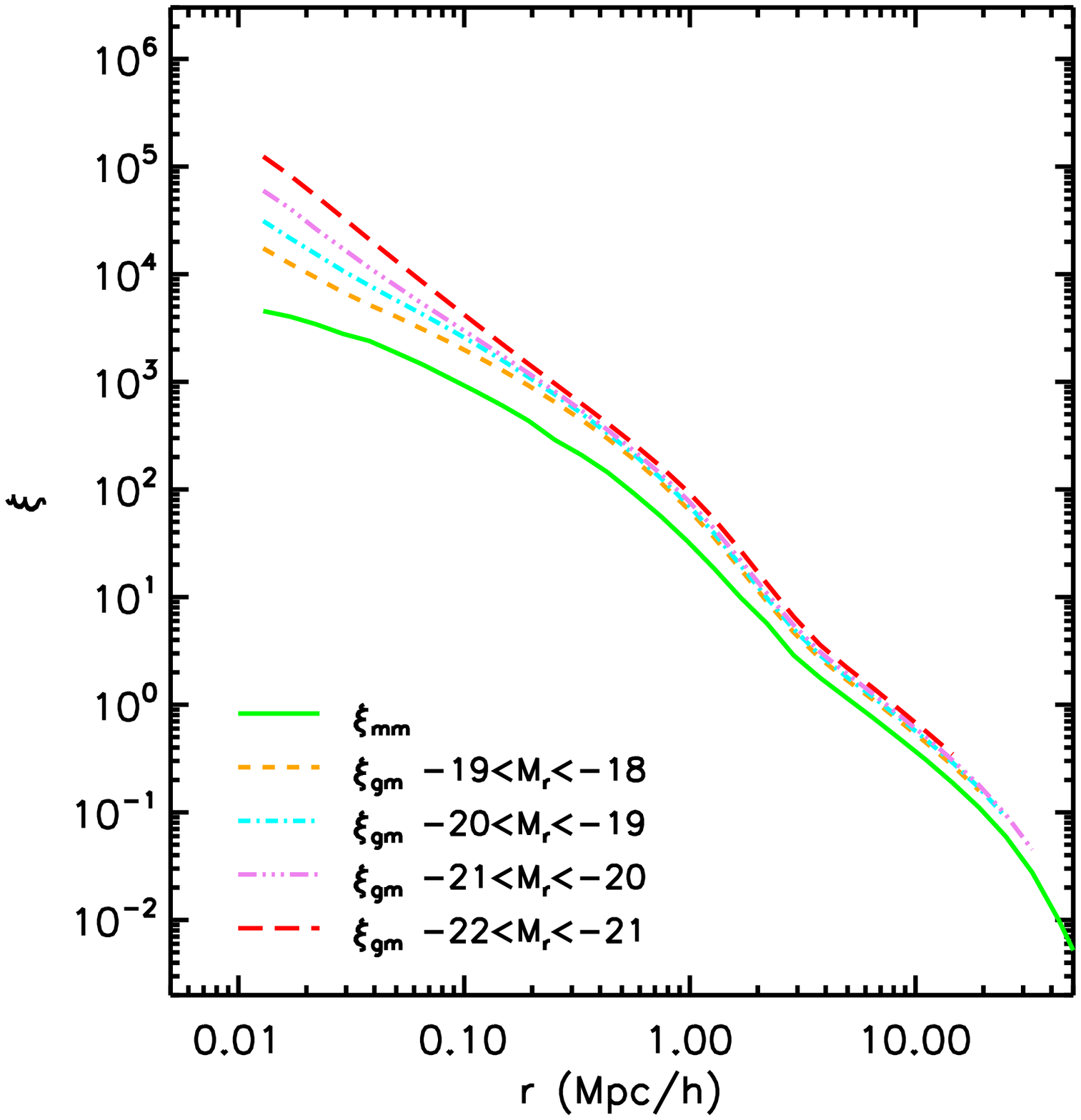}{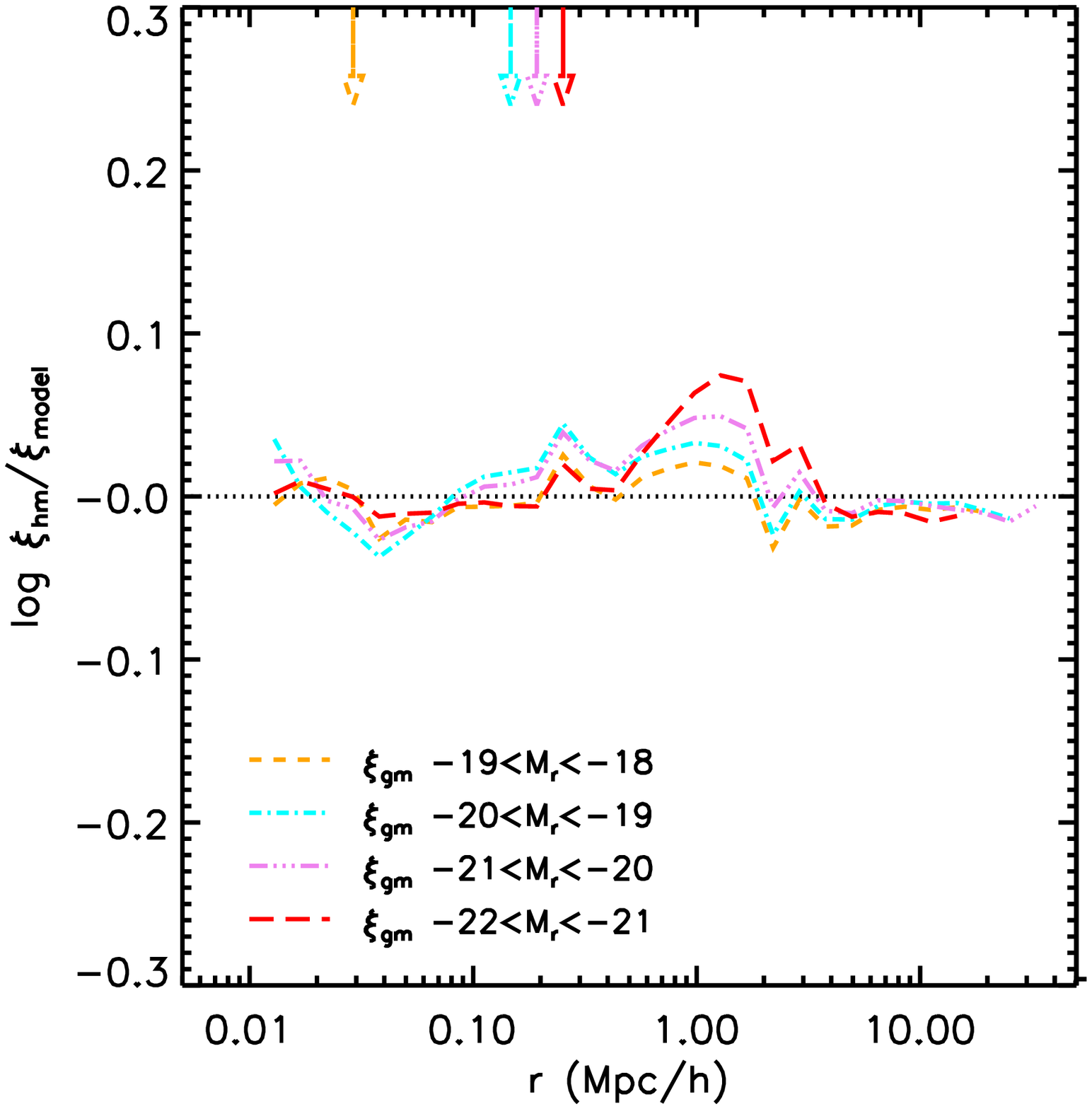}
\caption{ {\it Top panel:} Cross-correlations of the mass with {\it all}
satellite galaxies.  In this case the shape of $\xi_{\rm gm}$ is similar to
that of $\xi_{\rm mm}$ with an upturn on small scales due to the subhalo
mass associated with individual satellites that have not been fully tidally
disrupted.  {\it Bottom panel:} Deviations between the satellite $\xi_{\rm
gm}$ and the model given by eq.~\ref{eq:xisat}.  The deviations are largest at
intermediate scales, $r \simeq 1~\Mpch$, where the model fails to accurately
describe the shape of the scale-dependent bias.\label{fig:xigmsat}}
\end{figure}

The cross-correlation with mass for {\it all} satellite galaxies is shown in
Figure~\ref{fig:xigmsat}.  This is simply a linear combination of $\xi_{\rm
gm}$ for the subhalo and orphan satellites, weighted by the relative fractions
of each type. The shape of the resulting function appears quite simple: unlike
$\xi_{\rm gm}$ for the subhalo satellites, $\xi_{\rm gm}$ for the total
satellite sample is positively biased with respect to $\xi_{\rm mm}$ on all
scales, even for faint galaxy samples.  In fact, its shape follows that of
$\xi_{\rm mm}$ quite closely, with the addition of an upturn on small scales
due to the mass associated with the individual subhalos.  Encouraged by this,
we propose the following model for cross-correlations between satellite
galaxies and the mass:
\begin{equation}
\begin{split}
\xi_{\rm gm, sat}(r) = & \frac{\rho_{\rm halo}(r; c, M)}{\rhomean} + \\
&b(M_{\rm host})
\xi_{\rm mm}(r) \left[1+ \beta \exp\left(-\frac{r}{r_\beta}\right)\right].
\label{eq:xisat}
\end{split}
\end{equation}
On scales $r \gg r_\beta$, $\xi_{\rm sat}$ is equal to the product of the mass
autocorrelation function and the bias factor $b(M_{\rm host})$, where $M_{\rm
host}$ is the average mass of halos which host satellite galaxies of a given
luminosity.  On scales $r \lsim 2~\Mpch$, a scale-dependent bias is apparent in
the satellite $\xi_{\rm gm}$, which we attempt to model with an exponential
function.  In eq.~\ref{eq:xisat}, the parameters $r_\beta$ and $\beta$ control
the characteristic scale and amplitude of this scale-dependent bias.  On small
scales, $r \lsim 0.1~\Mpch$, the satellite model is dominated by the first term
due to the subhalo density profile.  

Figure~\ref{fig:xigmsat} shows the deviations of this model from the satellite
$\xi_{\rm gm}$ measured in the simulation.  Since a nonlinear model for
$\xi_{\rm mm}$ is used as the basis of the satellite model on large scales,
the systematic residuals in the fits to the halo and central galaxy
cross-correlation functions are not present in the satellite model fits.  The
main deficiency in the satellite model is in reproducing the scale-dependent
bias.  The fits shown in Figure~\ref{fig:xigmsat} were obtained using single
values of $r_\beta = 2~\Mpch$ and $\beta = 0.5$.  They could be improved
somewhat by varying these values for the different satellite galaxy luminosity
bins, but systematic residuals still remain because the exponential function
is only a crude match to the shape of the scale-dependent bias.

\begin{figure}
\plottwovert{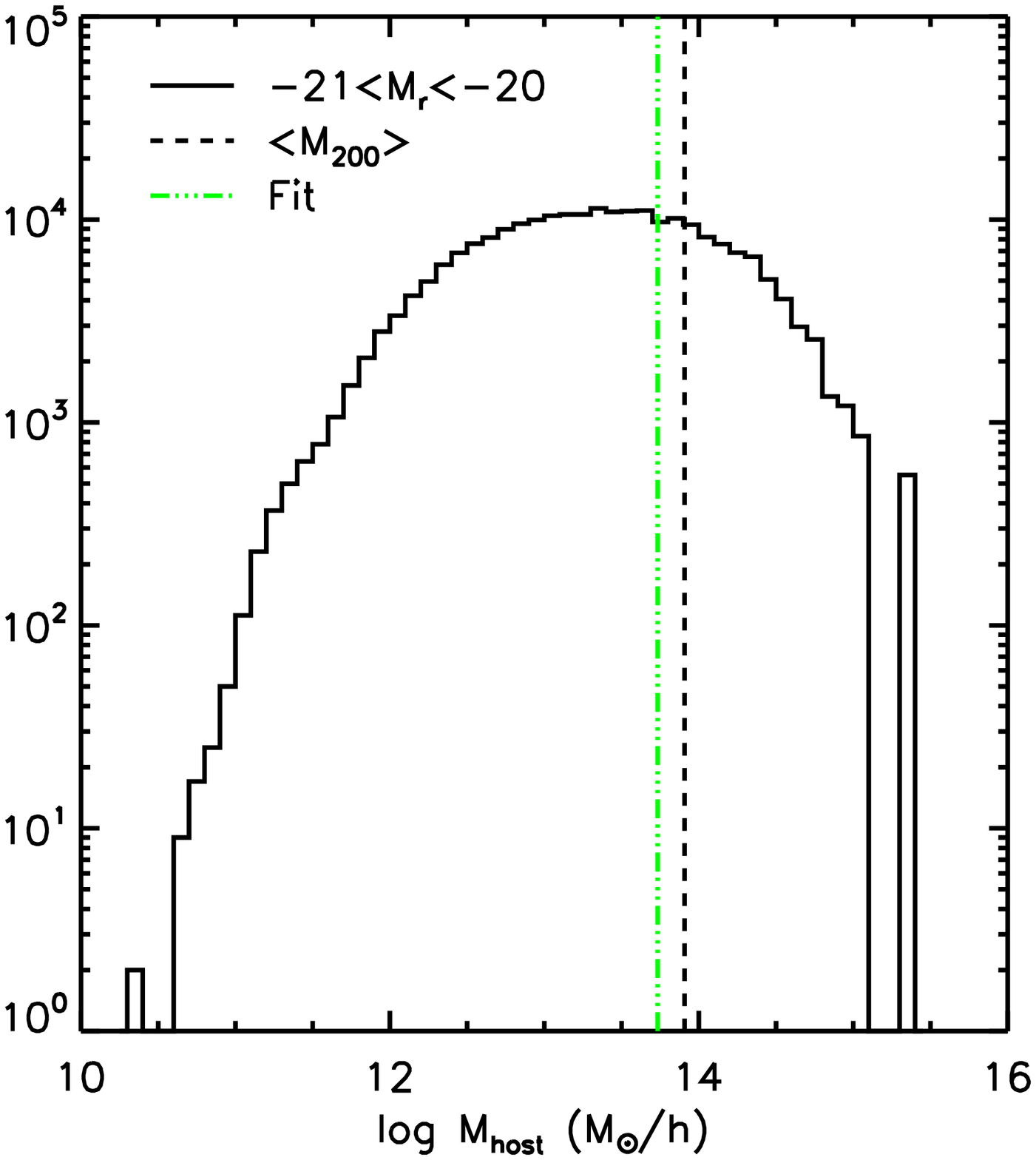}{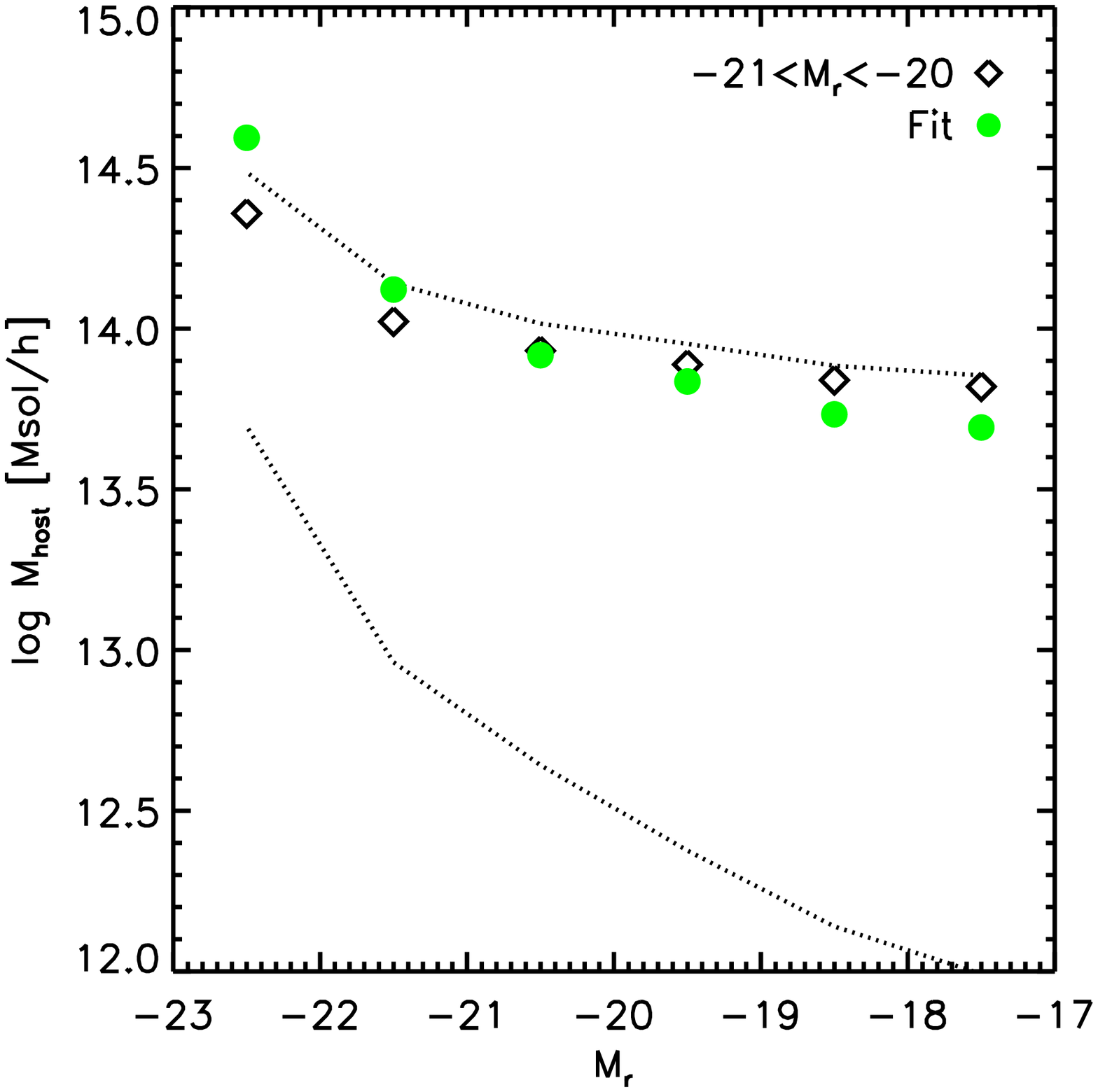}
\caption{ {\it Top panel:} Distribution of host halo $M_{200}$ for satellite
galaxies with $-21 < M_r < -20$.  {\it Bottom panel:} Mean host halo $M_{200}$
(open diamonds) with 20-and 80-percentile values (dotted lines) are compared
with best-fit values recovered by modelling the satellite-mass
cross-correlations (solid circles).
\label{fig:mhostdist}}
\end{figure}

The distribution of host halo masses for satellite galaxies in the luminosity
range $-21 < M_r < -20$ is shown in Figure~\ref{fig:mhostdist}.  The
distribution spans five orders of magnitude in mass. Nevertheless, the
best-fit model recovers the mean halo mass to within about $30\%$.  The
bottom panel shows mean and best-fit halo mass values as a function of $M_r$.
The values obtained from fitting the cross-correlations are accurate for host
halo masses in the range $-22 < M_r < -19$, but can differ from the true
values by as much as $50\%$ for host halos outside this intermediate range.

The model parameters entering the first term in eq.~\ref{eq:xisat} are the
concentration, $c$, and the mass, $M_{200}$, of the halo density profile
fitting formula.  However, the best fitting parameter values (shown in
Figure~\ref{fig:mhostdist}; the concentration is taken as the typical value
for halos of each mass) are not easily interpreted.

\subsection{Cross-correlations with mass for all galaxies}
\label{sec:allgal}

\begin{figure}
\plottwovert{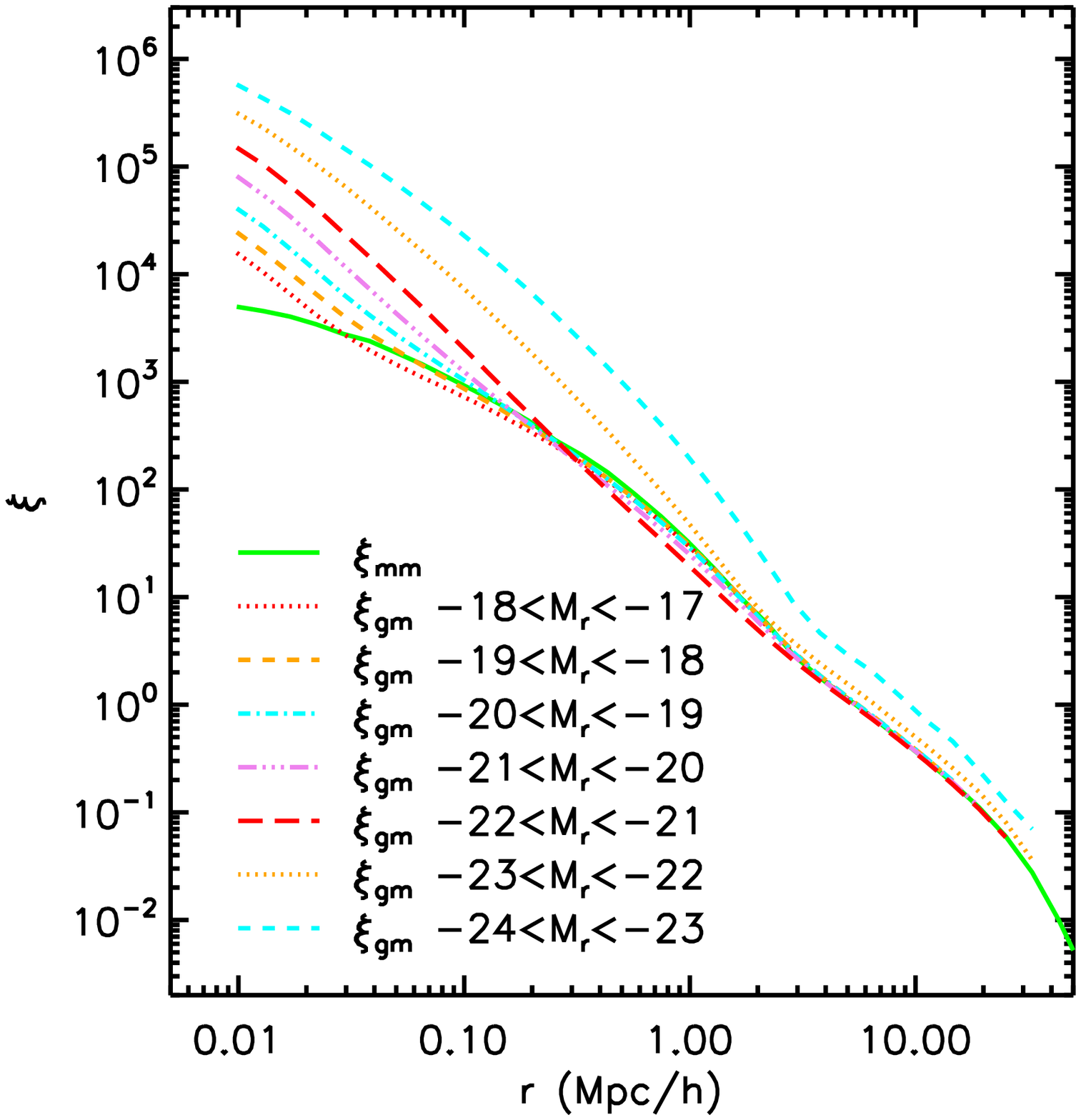}{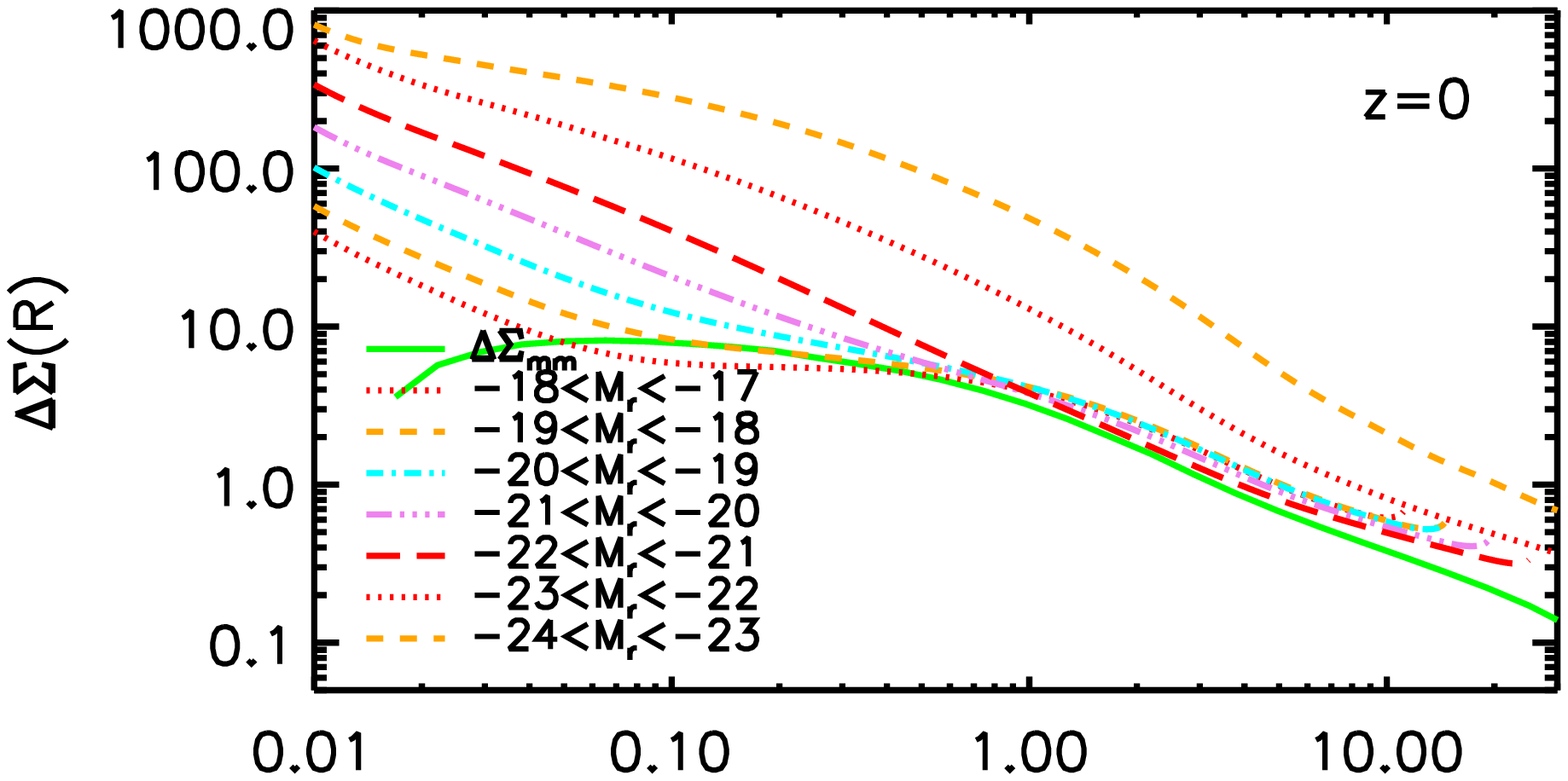}
\caption{ {\it Top panel:}  Cross-correlations between all galaxies and the
mass at $z=0$.  High-luminosity galaxy samples are dominated by central
galaxies, therefore $\xi_{\rm gm}$ resembles our simple model for $\xi_{\rm
hm}$.  Low-luminosity galaxy samples contain a significant fraction of
satellite galaxies and $\xi_{\rm gm}$ follows the shape of $\xi_{\rm
mm}$. {\it Bottom panel:} $\Delta \Sigma(R)$ corresponding to $\xi_{\rm gm}$.
\label{fig:xigmall}}
\end{figure}

In Figure~\ref{fig:xigmall} we present cross-correlations of all galaxies with
mass.  This is simply the linear combination of the cross-correlations for
central and satellite galaxies, weighted by the relative fractions of each
type, i.e.:
\begin{equation}
\xi_{\rm gm, all} = (1-f_{\rm sat}) \xi_{\rm gm, central} + f_{\rm sat} \xi_{\rm
gm, sat}, 
\label{eq:xigmall}
\end{equation}
where the fraction of satellite galaxies is defined as $f_{\rm sat} = N_{\rm
sat}/(N_{\rm central} + N_{\rm sat})$.  The satellite fraction in each
luminosity range strongly affects the shape of $\xi_{\rm gm}$ for galaxies of
that luminosity.  Table~\ref{tab:gals} shows that $f_{\rm sat}$ ranges from
$\simeq 40\%$ in the lowest luminosity bins, $M_r > -19$, to $\simeq 5\%$ for
$M_r < -22$.

Since eq.~\ref{eq:xigmall} is a linear combination of two functions which can
differ in value by several orders of magnitude and since $f_{\rm sat}$ is of order
unity, $\xi_{\rm gm, all}$ is dominated by the larger of $\xi_{\rm gm, central}$
and $\xi_{\rm gm, sat}$ at most radii.  For instance, for $-24 < M_r < -23$,
$\xi_{\rm gm, central}/\xi_{\rm gm, sat} \gg 1$ at all radii, thus
eq.~\ref{eq:xigmall} gives $\xi_{\rm gm, all} \simeq (1-f_{\rm sat})\xi_{\rm gm,
central}$ and the cross-correlation function is very similar to $\xi_{\rm gm,
central}$.  At lower luminosities, $\xi_{\rm gm, sat}/\xi_{\rm gm, central} \gg
1$ at intermediate radii, $0.1~\Mpch < r < 1~\Mpch$, so $\xi_{\rm gm, all}
\simeq f_{\rm sat}\xi_{\rm gm, sat} $ and in this case the cross-correlation
function resembles that of the satellite galaxies.

The combined $\xi_{\rm gm}$ model for central and satellite galaxies contains
the following five parameters: $M_{\rm central}, M_{\rm host}, M_{\rm sat},
c_{\rm sat},$ and $f_{\rm sat}$.  However, in fitting this composite model to
the measured cross-correlations, we obtained best-fit parameter values which
were not in good agreement with the true values of $M_{\rm central}, M_{\rm
host}$ and $f_{\rm sat}$.  This suggests that the shape of $\xi_{\rm gm}$ is
degenerate, with various combinations of the five free parameters giving almost
equivalent fits.

\begin{figure}
\plotone{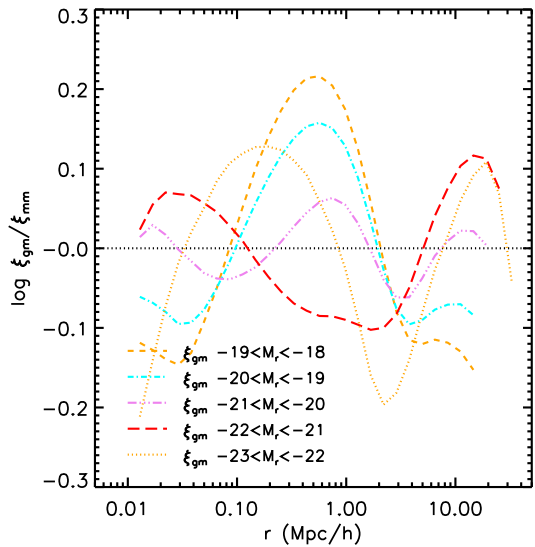}
\caption{Deviations from the best-fitting power laws of our measured
  cross-correlations between all galaxies and the mass.\label{fig:xigmdevlin}}
\end{figure}

It is interesting to note that our cross-correlation model is based on
components which deviate strongly from power laws, i.e., $\xi_{\rm lin}$,
$\xi_{\rm mm}$ and $\rho_{\rm halo}$. Nevertheless, at intermediate
luminosities $\xi_{\rm gm, all}$ is reasonably well described by a power-law
over a wide range in radius.  In order to illustrate this point, we plot in
Figure~\ref{fig:xigmdevlin} the deviations from power-law fits to 
$\xi_{\rm gm, all}$.  In particular, we find that in the range $-21 < M_r <
-20$, $\xi_{\rm gm, all}$ is well fit by a power law with slope $r^{-1.8}$ and
the deviations from the fit are $\lsim 10\%$ at all radii!

For galaxies more luminous than $M_r < -21$, the deviations from best-fit
power laws become $\gsim 50\%$ at some radii.  Galaxies in this luminosity
range are dominated by central galaxies, so our model for $\xi_{\rm gm,
central}$ can be used to fit observations and provide detailed checks on
various aspects of the cosmological and galaxy formation models.  At lower
luminosities, however, we conclude that some independent information regarding
the satellite fraction must be included in order to extract useful information
about the average mass of the halos which host satellites. Alternatively,
shear data can be stacked around galaxies that are observed to be brighter
than all their neighbours, and thus are very likely central galaxies. Our
models can be used to predict $\Delta \Sigma(R)$ for direct comparison with
such galaxy-galaxy lensing measurements.  Examples are shown in the lower
panel of Figure~\ref{fig:xigmall}. It is notable that these curves show a
variety of scales and features which should be directly observable.

\section{Summary}
\label{sec:summary}

We have calculated cross-correlations between halo centres and the mass, and
between galaxies and the mass in the Millennium Simulation of a \LCDM
cosmology.  The shape of the halo-mass cross-correlation function $\xi_{\rm
hm}$ is well fit by a simple two-part model.  On small scales, $\xi_{\rm hm}$
is specified by the halo density profile which is accurately described by the
Einasto model advocated by N04.  On large scales, $\xi_{\rm hm}$ is specified
by the mass autocorrelation function predicted by linear theory and the halo
bias model of \cite{SHETH99}.  The deviations between $\xi_{\rm hm}$ and the
best-fit model are dominated by the quasi-linear distortion in $\xi_{\rm mm}$,
but are otherwise $\lsim 5\%$.

The cross-correlation functions of central galaxies, $\xi_{\rm gm, central}$,
are reasonably well fit by our model for $\xi_{\rm hm}$. The best-fit models
recover the mean halo masses of central galaxies to within $30\%$.  The
cross-correlations of satellite galaxies, $\xi_{\rm gm, sat}$ appear
qualitatively different from those of halos or central galaxies.  Their shape
is similar to that of $\xi_{\rm mm}$, with an upturn at small scales due to
the mass associated with the individual subhalos in which most satellites
reside.  A model for $\xi_{\rm gm, sat}$ based on these features reproduces
the cross-correlation functions to within $10\%$ and recovers the mean host
halo mass to within $50\%$.

The cross-correlation function of all galaxies, $\xi_{\rm gm, all}$ is simply
a linear combination of $\xi_{\rm gm, central}$ and $\xi_{\rm gm, sat}$
weighted by the relative fractions of central and satellite galaxies.  For
very luminous galaxies, the satellite fraction $f_{\rm sat} \lsim 10\%$, and
$\xi_{\rm gm, all}$ is dominated by the contribution of $\xi_{\rm gm,
central}$ . At intermediate luminosities, $-22 < M_r < -20$, $\xi_{\rm gm,
all}$ is reasonably well fit by a power law. At lower luminosities the
cross-correlation is dominated by the satellite galaxy contribution.

The conversion from the three-dimensional cross-correlations to the directly
observable mean tangential shear (which is proportional to $\Delta\Sigma(R)$,
the difference between the mean enclosed surface density and the local surface
density at each projected radius $R$) accentuates features in the
cross-correlations. Galaxy-galaxy lensing surveys typically contain enough
information to separate the lenses into likely satellites and likely central
systems. As a result, the features seen in our predictions should provide
information on cosmological parameters, tidal stripping processes, and the exact
way in which galaxies trace the dark matter. If the features are clearly
detected where they are expected, this will provide a major challenge to
theories which try to replace dark matter by a modification of Einsteinian
gravity.

\section*{Acknowledgments}
\label{acknowledgements}
We thank L.~Gao, U.~Seljak, N.~Padmanabhan, A.~Leauthaud, and R.~Mandelbaum for
useful discussions.

\bibliographystyle{../Halos2/astron}
\bibliography{../stan}

\end{document}